\documentclass[%
 reprint,
 superscriptaddress,
nofootinbib,
 amsmath,amssymb,
 aps,
 pra,
]{revtex4-2}

\usepackage{graphicx}% Include figure files
\usepackage{dcolumn}% Align table columns on decimal point
\usepackage{bm}% bold math  
\usepackage{url}
\usepackage[colorlinks,citecolor=blue,linkcolor=blue,urlcolor=blue]{hyperref}
\usepackage{enumitem}
\usepackage{framed}
\usepackage{xcolor}

\usepackage{mathtools}
\def\f{\frac}
\def\k{\kappa}

\def\be{\beta}
\def\D{\Delta}

\def\t{\tau}
\def\a{\alpha}
\def\mP{\mathcal{P}}
\def\mI{\mathcal{I}}
\def\mS{\mathcal{S}}
\def\mJ{\mathcal{J}}
\def\s{\sigma}
\def\kb{k_B}
\def\d{\delta}
\def\p{\partial}
\def\l{\lambda}
\def\L{\Lambda}
\def\G{\Gamma}
\def\o{\Omega}

\def\g{\gamma}
\def\jv{ {\bf j}}
\def\jr{ {\bf j}_r}
\def\jd{ {\bf j}_d}
\def\jdd{ { j}_d}
\def\jrr{ { j}_r}

\def\zrr{\mathbf{u_+}}
\def\zii{\mathbf{u_-}}

\newcommand{\avg}[1]{\langle #1 \rangle}
\newcommand{\Bavg}[1]{\Big\langle #1 \Big\rangle}

\newcommand{\llangle}{\langle\!\langle}
\newcommand{\rrangle}{\rangle\!\rangle}
\newcommand{\avgg}[1]{\llangle #1 \rrangle}

\begin{document}

\title{A variational formulation of stochastic thermodynamics: Finite-dimensional systems}

\author{H\'ector Vaquero del Pino}
\email{hector001@e.ntu.edu.sg}
\affiliation{Division of Physics and Applied Physics, School of Physical and Mathematical Sciences,
Nanyang Technological University, 21 Nanyang Link, Singapore 637371}

\author{Fran\c{c}ois Gay-Balmaz}
\email{francois.gb@ntu.edu.sg}
\affiliation{
 Division of Mathematical Sciences, School of Physical and Mathematical Sciences,
Nanyang Technological University, 21 Nanyang Link, Singapore 637371
}

\author{Hiroaki Yoshimura}
\affiliation{School of Science and Engineering, Waseda University. 3–4–1, Okubo, Shinjuku, Tokyo 169-8555, Japan}

\author{Lock Yue Chew}
\email{lockyue@ntu.edu.sg}
\affiliation{Division of Physics and Applied Physics, School of Physical and Mathematical Sciences,
Nanyang Technological University, 21 Nanyang Link, Singapore 637371}

\date{\today}% It is always \today, today,

\begin{abstract}
In this paper, we develop a variational foundation for stochastic thermodynamics of finite-dimensional, continuous-time systems. Requiring the second law (non-negative average total entropy production) systematically yields a consistent thermodynamic structure, from which novel generalized fluctuation–dissipation relations emerge naturally, ensuring local detailed balance. This principle extends key results of stochastic thermodynamics including an individual trajectory level description of both configurational and thermal variables and fluctuation theorems in an extended thermodynamic phase space. It applies to both closed and open systems, while accommodating state-dependent parameters, nonlinear couplings between configurational and thermal degrees of freedom, and cross-correlated noise consistent with Onsager symmetry. This is achieved by establishing a unified geometric framework in which stochastic thermodynamics emerges from a generalized Lagrange–d’Alembert principle, building on the variational structure introduced by Gay-Balmaz and Yoshimura [\href{https://doi.org/10.1098/rsta.2022.0280}{Phil. Trans. R. Soc. A \textbf{381}, 2256 (2023)}]. Irreversible and stochastic forces are incorporated through nonlinear nonholonomic constraints, with entropy treated as an independent dynamical variable. This work provides a novel approach for thermodynamically consistent modeling of stochastic systems, and paves the way to applications in continuum systems such as active and complex fluids.
\end{abstract}

\maketitle

%\tableofcontents

\section{\label{sec:intro} Introduction}
Stochastic thermodynamics (ST) is the framework developed to systematically understand the thermodynamics associated with systems where fluctuations play an important physical role in the dynamics, e.g. mesoscopic systems and critical phenomena. The need for this formalism is motivated especially for nonequilibrium systems, where the conventional tools of statistical mechanics do not apply. This framework builds on the extensions of the first and second laws to the stochastic regime, providing a description at the individual trajectory level, as well as universal constraints on the distributions that hold beyond linear response theory. Under the assumption of time-scale separation, the observable (slow) and thermal (fast) degrees of freedom (DoFs) are typically modeled by Markovian dynamics, such as master equations (for discrete-time systems) or Langevin equations (for continuous-time systems) \cite{Seifert_2012}. 

However, the extensions of the first and second laws to the stochastic regime in ST are not systematically connected. On one hand, entropy acquires a new definition from the perspective of information theory as a raw measure of irreversibility \cite{pelitti,VANDENBROECK20156,Seifert_2005,Seifert_2012,ActiveFields,Fodor_2022}, while stochastic energetics is approached from a different point of view, namely Sekimoto's approach \cite{Sekimoto2010}. To reconcile the stochastic first and second laws, we require our model to satisfy thermodynamic consistency, providing a well-defined physical interpretation of entropy beyond its information-theoretic meaning. The principle of local detailed balance (LDB) offers precisely this constructive link, which fundamentally stems from the time-reversal symmetry of the microscopic laws of Physics or micro-reversibility, see Ref.~\cite{Maes_LDB}.  

For continuous-time systems, which are the focus of this paper, the principle of LDB manifests in the Langevin equation through the fluctuation-dissipation relation (FDR), in order to ensure thermodynamic consistency. We emphasize that throughout this work we refer to FDR as generalized fluctuation-dissipation Einstein relation linking noise and dissipation \cite{Broeck_1,Broeck_2,Polettini, VANDENBROECK20156}, also termed FDR of the second kind, and is not to be identified with the FDR of the first kind as in how the system responds to an external perturbation \cite{Maes_FDR}. Note that imposing LDB in the set-up of Markov dynamics implies an FDR of the second kind, which in turn implies a standard FDR of the first kind \textit{near} equilibrium, see Ref.~\cite{Maes_FDR}. 

Thus, the principle of LDB is essential for constructing physically meaningful and thermodynamically interpretable models \cite{ActiveFields,PhysRevX.15.021050,Maes_LDB}. While there are different methods to derive FDRs, including fixing the equilibrium distribution in the Fokker-Planck steady state \cite{zwanzig, Kardar2007} or imposing the LDB condition directly \cite{Maes_LDB,ActiveFields}, determining such relations in a general setting in the context of Langevin dynamics remains challenging, and the Einstein relation \cite{Einstein1905} is often assumed by default \cite{Seifert_2012}. As a result, the model at hand often lacks a clear thermodynamic interpretation \cite{Maes_LDB}. The goal of this paper is to present a novel general and systematic approach to the thermodynamically consistent modeling of stochastic finite-dimensional, Langevin systems. To this end, we extend the variational formulation of nonequilibrium thermodynamics presented in Ref. \cite{GBYo2019} to the stochastic regime, thereby establishing a principled connection between ST and analytical mechanics, in line with the growing interest \cite{Beyen_2025}.

The variational approach in Ref. \cite{GBYo2019} generalizes classical mechanics to include irreversible processes by extending the Lagrange-d'Alembert principle to incorporate entropy production as a nonlinear nonholonomic constraint, with phenomenological laws embedded directly into the constraint structure. A key insight of this framework is the introduction of thermodynamic displacements, conjugates to the thermal variables, which provide a natural definition of the variational constraint and generalize earlier notions of thermal displacement. The resulting variational structure yields a complete set of evolution equations, formulated as a system of ordinary differential equations (ODEs) for discrete systems, enabling the determination of all system variables at all times. It is important to note that this variational formulation consistently recovers Hamilton’s principle in the absence of irreversible processes, and---as a macroscopic description---assumes local thermodynamic equilibrium.

By extending this framework to the stochastic regime, we generalize it to describe systems driven far from equilibrium. The proposed variational principle for ST is built from the first and second laws of thermodynamics, under the assumptions of time-scale separation, large bath size and micro-reversibility. The FDR can be derived systematically in general settings without assuming a fixed steady state, without relying on problematic path probabilities, and without requiring thermal baths to be infinite. Instead, results from statistical mechanics are recovered in the appropriate limits, while the formulation reproduces macroscopic nonequilibrium thermodynamics in the vanishing-noise limit. Moreover, as shown in subsequent sections, this approach naturally extends key features of ST, including an individual trajectory level description of both configurational and thermal variables, and fluctuation theorems (FTs) in the thermodynamic phase space. Remarkably, once thermodynamic consistency is enforced via the FDRs, the equations for entropy production derived from the variational approach assume the same form as in standard ST, up to a reformulation in an extended phase space.

Here two remarks are in order. Firstly, in Ref. \cite{Maes_FDR} it is illustrated how not in all physical cases LDB needs to be true, providing examples such as non-equilibrium baths
or large system-bath/bath-bath couplings. The role of the variational approach in this regard will be assessed in Sec. \ref{sec:discussion}. Secondly, as the variational principle is framed in Lagrangian mechanics, it only covers Langevin-type dynamics and not the more general setting of Markov chains.

The paper is organized as follows. Section \ref{sec:ST} provides an overview of the framework of ST, which can be skipped by readers already familiar with the
topic, except perhaps for subsection \ref{sec:ST_caveats}. Section \ref{sec:VP} introduces the variational principle for ST and illustrates it with examples of (finite-dimensional) closed and open systems. Section \ref{sec:discussion} offers a discussion of our findings, and Section \ref{sec:conclusion} summarizes the main conclusions of this study. Additional analytical derivations are organized into six appendices at the end.

\section{\label{sec:ST} Stochastic thermodynamics}

\subsection{\label{sec:representations} Representations of Langevin processes}
The stochastic dynamics of continuous-time systems have three different but equivalent descriptions \cite{Seifert_2012,Sekimoto2010}: the Langevin equation, the Fokker-Planck equation and the path integral formulation.

In the general case of a $d$-dimensional particle system, the Langevin equation for underdamped dynamics can be written as $(\nu=1,...,d)$ \cite{Graham2}:
\begin{subequations}
\begin{align}
    &m\dot{x}^\nu = p^\nu, \label{eq:momentum}\\
    & \dot{p}^\nu = F^\nu(\mathbf{x},\mathbf{p},\lambda) + g^\nu _k(\mathbf{x},\mathbf{p};t)\circ\zeta^k (t),\label{eq:LE}
\end{align}
\end{subequations}
where $(\mathbf{x},\mathbf{p})$ correspond to position and momentum respectively, $F^\nu$ is a generalized force, $\lambda (t)$ denotes a time-dependent driving, and  $\zeta^\nu$ is a gaussian white noise verifying  
\begin{subequations}\label{eq:gaussian_noise}
\begin{eqnarray}
    &&\avg{\zeta^k(t)}=0,\\
    &&\avg{\zeta^k(t)\zeta^\ell(t') }=2\d^{k\ell}\d(t-t').
\end{eqnarray}
\end{subequations}
The product $\circ$ denotes the Stratonovich convention (SC), which will be applied throughout this paper due to its suitability for modeling physical processes \cite{Sekimoto2010}. Repeated indices are summed over according to Einstein's convention. The noise amplitude contribution $g(\mathbf{x},\mathbf{p};t)$ allows for the possibility of state-dependent noise. The corresponding Fokker-Planck equation (FP) is \cite{gardiner1985handbook}: 
\begin{align}\label{eq:FP}
  \partial_t \mP &= -\partial_{x^\nu} \left(\frac{p^\nu}{m}\mP\right)
  -\partial_{p^\nu}\left[\left(F^\nu -g^\nu_k \partial_{p^\mu}g^\mu_k\right)\mP
  -g^\nu_k g^\mu_k \partial_{p^\mu}\mP\right] \nonumber\\
  &= -\nabla_{\mathbf{x}}\cdot\mathbf{j}_{x}-\nabla_{\mathbf{p}}\cdot\mathbf{j}_{p} ,
\end{align}
where $\mP(\mathbf{x},\mathbf{p},t)$ denotes the probability density function (PDF) of a certain configuration $(\mathbf{x},\mathbf{p})$ at time $t$, and $\mathbf{j}$ the probability current. While the Langevin equation \eqref{eq:momentum}-\eqref{eq:LE} provides dynamical information of an individual trajectory, the FP equation \eqref{eq:FP} provides the ensemble level information. 

An additional representation which is applied in ST is the path integral (PI) formulation. Similarly as for the Langevin equation, this description allows assigning thermodynamic observables to individual trajectories. Such path integral is the analog of the propagator $K$ in Quantum Mechanics, where the analog of the Hamiltonian would in this case correspond to the Fokker-Planck operator \cite{Cugliandolo_2019}:
\begin{subequations}
\begin{eqnarray}
     &&\mP(\mathbf{z},t)=\int _{\mathbb{R}^{2d}}d\mathbf{z}_0 \, K(\mathbf{z},t;\mathbf{z}_0,t_0) \mP_0(\mathbf{z}_0),\label{eq:propagator}\\
    &&K(\mathbf{z}_t,t;\mathbf{z}_0,t_0) = \int_{\mathbf{z}_0}^{\mathbf{z}_t} \mathcal{D}\mathbf{z}\,P[\mathbf{z}|\mathbf{z}_0] . 
\end{eqnarray}
\end{subequations}
For convenience, we denote the full phase-space variable as $\mathbf{z} = (\mathbf{x}, \mathbf{p})$. The initial and final times are denoted by $t_0$ and $t$, respectively, with corresponding configurations $\mathbf{z}_0$ and $\mathbf{z}_t$. The path probability can be defined as
\begin{equation}
   P[\mathbf{z}|\mathbf{z}_0]=\mathcal{N}\{\mathbf{z}_n\}\,  \exp (-\mathcal{A}[\mathbf{z}]), 
\end{equation}
where $\mathcal{A}[\mathbf{z}]$ is the action functional and $\mathcal{N}\{\mathbf{z}_n\}$ the normalization factor defined for a discretized temporal grid $\{t_n = \epsilon n\}_{n=0}^N$. Since to each Langevin equation corresponds a unique FP equation, the PDF $\mP(\mathbf{z},t)$ is uniquely defined for a given interpretation (Stratonovich/Ito). However, this only uniquely defines $K(\mathbf{z}_t,t;\mathbf{z}_0,t_0)$ in the continuous limit,  allowing the freedom to define different normalizations $\mathcal{N}\{\mathbf{z}_n\}$ or actions $\mathcal{A}[\mathbf{z}]$ as long as Eq. \eqref{eq:propagator} is fulfilled \cite{Wissel1979}. 

As a result, different expressions for the path integral (PI) corresponding to the same interpretation of the noise can be found in the literature \cite{Onsager_Machlup,Lau_2007,Cates_2022, Cugliandolo_2017, Cugliandolo_2019, Graham1977, Graham2, Deininghaus1979}. However, this ambiguity disappears in the context of ST when the contribution of the normalization $\mathcal{N}\{\mathbf{z}_n\}$ is properly taken into account. In this case, the results at the level of individual trajectories obtained via the PI formalism coincide for its different forms, as well as with those derived from the FP formalism---which is uniquely defined---when ensemble averages are computed, as required for consistency.

A convenient expression for the PI in the context of ST is the Onsager-Machlup (OM) action \cite{Onsager_Machlup,Lau_2007,Cates_2022}. For a discretized temporal grid $\{t_n = \epsilon n\}_{n=0}^N$, the propagator and the action\footnote{Note the action is written in the continuous time limit.} are given, respectively, by:
\begin{equation}\label{eq:PI_multid}
\begin{aligned}
      & K(\mathbf{z}_t,t;\mathbf{z}_0,t_0) = \int_{\mathbf{z}_0}^{\mathbf{z}_t}\mathcal{D}\mathbf{z}\prod_{n=0}^{N-1}\f{1}{\det(g^{\nu\mu}_{n+1/2})(4\pi\epsilon)^{d/2}}\, e^{-\mathcal{A}[\mathbf{z}]},\\
        &\mathcal{A}[\mathbf{z}] =\int_{t_0}^{t}d\tau \left[ \f{1}{2}\partial_{p^\mu}  F^\mu  +\f{1}{4}\left(\partial_{p_\nu} g^\mu_i \partial_{p_\mu} g^\nu_i -\partial_{p_\mu} g^\mu_i \partial_{p_\nu} g^\nu_i \right)\right.\\
        &\left. \qquad + \f{1}{4}(\dot{p}^\mu-F^\mu + g_{i}^\mu\partial_{p^\a}  g_{i}^\a)D_{\mu\nu}(\dot{p}^\nu-F^\nu + g_{i}^\nu\partial_{p^\a}  g_{i}^\a)\right],
\end{aligned}
\end{equation} %dz_n^\nu = \prod_\mu dp_n^\nu dx_n^\mu
with  $D^{\nu\mu}= g_{i}^\nu g_{i}^\mu$ denoting the covariance matrix and $D_{\mu\nu}=(D^{\nu\mu})^{-1}$ its inverse (provided it exists), such that $D_{\mu\nu}D^{\nu\a}=\d_\mu^\a$. 

We have already accounted for  Eq. \eqref{eq:momentum} via the Dirac delta term, and we denote the constrained integral over paths as 
\begin{equation}
    \mathcal{D}\mathbf{z}=\mathcal{D}\mathbf{x}\mathcal{D}\mathbf{p}\d^{(d)} (\mathbf{p}-m\mathbf{\dot x})=\lim_{\epsilon\rightarrow 0} \prod_{n=1}^{N-1}\prod_\nu dz_n^\nu  \d(m\dot{x}^\nu - p^\nu).
\end{equation}
This contribution can be understood as the $D\rightarrow 0$ limit of $\mathbf{p}=m \mathbf{\dot x} + \bm{\xi}$ with $\avg{\xi^i(t)\xi^j(t')}=2D\d^{ij}\d(t-t')$. Due to addition of actions, the PI over a single time step includes the contribution:
\begin{eqnarray}
     P(\mathbf{z}_{n+1} | \mathbf{z}_n)&\propto& \lim_{D\rightarrow0}\f{1}{(4\pi D\varepsilon)^{d/2}}\exp \left\{ -\varepsilon\f{||m \mathbf{\dot x} - \mathbf{p}||^2}{4D}\right\} \nonumber\\
     &=&\d ^{(d)}(\mathbf{p}-m\mathbf{\dot x}),
\end{eqnarray}
where the last equality corresponds to the definition of the Dirac delta function as a limit distribution \cite{Arfken}.

The advantage of the OM form lies in the fact that its associated measure is symmetric under time reversal. This symmetry arises because the points in the time-reversed trajectory, defined by $\hat{x}_{n+a} = x_{n+(1-a)}$, coincide with those in the forward trajectory when using the SC, i.e. $a = 1/2$. As a result, only the terms in the action contribute to the entropy production. In contrast, other expressions of the path integral---such as those in Refs. \cite{Graham1977, Cugliandolo_2019}---involve measures that are not symmetric under time reversal, either due to asymmetric discretization schemes or the presence of terms that are odd under time reversal. These asymmetries must therefore be explicitly accounted for when computing entropy production to avoid inconsistencies \cite{Cates_2022}. However, we emphasize that these alternative formulations yield covariant actions, in contrast to the OM form \cite{Cugliandolo_2019}.

\subsection{\label{sec:stoch_entropy_prod} Stochastic entropy production}
In ST, entropy production (EP) is defined as the breaking of time-reversal symmetry (TRSB), and is quantitatively measured by the Kullback-Leibler (KL) divergence---or statistical distance---between the probability distributions of a forward process and its time-reversed counterpart \cite{Parrondo_2009, pelitti, VANDENBROECK20156}:
\begin{equation}\label{eq:KL}
    \langle \Delta s_{\rm tot} \rangle := D_{KL}[P[\mathbf{z}]||P[\hat{\mathbf{z}}]]=\int_{\mathbf{z}_0}^{\mathbf{z}_t} \mathcal{D}\mathbf{z} P[\mathbf{z}]\text{ln}\frac{P[\mathbf{z}]}{P[\hat{\mathbf{z}}]},
\end{equation}
where---setting $t_0 = 0$ without loss of generality---$\mathbf{z}(\t)=(\mathbf{x}(\t),\mathbf{p}(\t))$  denotes the forward trajectory, and $\hat{\mathbf{z}}(\t)=(\mathbf{x}(t-\t),\mathbf{p}(t-\t))$ the backward or time-reversed trajectory. In the presence of an externally controlled, time-dependent protocol $\lambda(\t)$, which reverses as $\hat{\lambda}(\t) = \lambda(t-\t)$, the path probability is then given by $P[\mathbf{z},\l]$ with reverse $P[\hat{\mathbf{z}},\hat{\l}]$.  Odd variables are those that change sign under time reversal, such as momentum or velocity, while even variables remain unchanged. 

The KL divergence is an invariant (under non-singular transformations), nonnegative flat operator \cite{KL}, such that the definition \eqref{eq:KL} already incorporates the second law of ST $ \langle \Delta s_{\rm tot} \rangle \geq 0$. Moreover, as a consequence of this definition, the integral fluctuation theorem (IFT) follows \cite{Seifert_2005}:
\begin{equation}\label{eq:IFT}
    \langle e^{-\Delta s_{\rm tot}}\rangle = \int \mathcal{D}\mathbf{z} P[\mathbf{z}]\frac{P[\hat{\mathbf{z}}]}{P[\mathbf{z}]}= \int  \mathcal{D}\mathbf{z} P[\hat{\mathbf{z}}] = 1
\end{equation}
where the last equality is drawn from the normalization condition and the fact that summing over all paths do not distinguish between forward and backward trajectories \cite{Seifert_2012}. 

The EP for an individual trajectory is obtained by undoing the path average:
\begin{subequations}
\begin{align}
    &\Delta s_{\rm tot} =\ln\frac{P[\mathbf{z}]}{P[\hat{\mathbf{z}}]} = \Delta s_m + \Delta s, \label{eq:Delta_s_tot}\\
    &\Delta s_m  :=\ln\frac{P[\mathbf{z}|\mathbf{z}_0 ]}{P[\hat{\mathbf{z}}|\mathbf{z}_t ]},\quad  \Delta s:=\ln\frac{\mathcal{P}(\mathbf{z}_0, t_0)}{\mathcal{P}(\mathbf{z}_t,t)} . \label{eq:EP}
\end{align}
\end{subequations}
The first term in Eq. \eqref{eq:EP} represents the stochastic EP in the medium $(\Delta s_m )$, while the second term accounts for the stochastic entropy change of the system $(\Delta s )$. The probability distributions $\mathcal{P}$ are governed by the corresponding Fokker-Planck equation. Upon taking the ensemble average of the system contribution, one recovers the Shannon entropy $\mS$ \cite{Seifert_2012}:
\begin{subequations}\label{eq:stoch_shannon}
    \begin{eqnarray}
        s(t)&&= - \ln \mathcal{P}(\mathbf{z}(t),t),\\
        \mS(t) &&:= \avg{s(t)} = -\int_{\mathbb{R}^{2d}}  d\mathbf{z}\,\mP(\mathbf{z},t)\ln \mP(\mathbf{z},t) .
    \end{eqnarray}
\end{subequations}

Taking the time derivative of the total entropy production yields the stochastic entropy production rate (EPR):
\begin{eqnarray} \label{eq:dot_s_tot}
    \dot s_{\rm tot}(t) &=& \dot s (t) + \dot s_m (t).
\end{eqnarray}
We illustrate the EPR by applying the framework to Eq. \eqref{eq:LE}. First, we define even and odd contributions to the force $F^\nu_\pm=(F^\nu\pm \hat F^\nu)/2$, where the time-reversed force is defined as $\hat{F}^\nu =F^\nu(\mathbf{x}(u),-\mathbf{p}(u),\l (u))$ for $u=t-\tau$. Then, we can identify the reversible and dissipative contributions to the probability current along the momentum dimension:
\begin{subequations}
\begin{eqnarray}
    && j_p^\nu = j^\nu_r + j^\nu_d ,\\
   && j^\nu_r = F^\nu_+ \mathcal{P} ,\\
   && j^\nu_d = (F^\nu_- -g^\nu_k \partial_{p_\mu}g^\mu_k)\mathcal{P}-D^{\nu\mu} \partial_{p_\mu}\mP.
\end{eqnarray}
\end{subequations}
The system EPR yields:
\begin{equation}\label{eq:stoch_s}
\begin{aligned}
    \dot{s}&=-\f{d}{dt}\ln \mP = -\f{\partial_t \mP}{\mP}-\f{\partial_{x_\mu} \mP}{\mP}\f{p^\mu}{m}-\f{\partial_{p_\mu} \mP}{\mP}\dot{p}^\mu \\
    &=  -\f{\partial_t \mP}{\mP}-\f{\partial_{x_\mu} \mP}{\mP}\f{p^\mu}{m}+ \dot{p}^\mu D_{\mu\nu}\left( \f{j^\nu_d}{\mP}-(F^\nu_- -g^\nu_k \partial_{p_\a}g^\a_k)\right)   ,
\end{aligned}
\end{equation}
where we applied 
\begin{equation}
    \partial_{p_\mu}\mP= D_{\mu\nu} \left((F^\nu_- -g^\nu_k \partial_{p_\a}g^\a_k)\mathcal{P}-j^\nu_d\right).
\end{equation}
The medium EP is derived by applying the PI \eqref{eq:PI_multid} in Eq. \eqref{eq:EP}:
\begin{equation}
\begin{aligned}
     \Delta s_m &=\ln\frac{P[\mathbf{z}|\mathbf{z}_0 ]}{P[\hat{\mathbf{z}}|\mathbf{z}_t ]}=\text{ln}\frac{\mathcal{N}\{\mathbf{z}_n\}\, \exp\{-\mathcal{A}\}}{\mathcal{N}\{\mathbf{\hat z}_n\}\, \exp\{-\hat{\mathcal{A}}\}}=\hat{\mathcal{A}}-\mathcal{A} \\
     &= \int_{t_0}^t d\tau \left( (\dot{p}^\mu - F^\mu_+) D_{\mu\nu} (F^\nu_- -g^\nu_k \partial_{p_\a}g^\a_k)-\partial_{p_\mu}F^\mu_+\right).
\end{aligned}
\end{equation}
Here, the backward action is given by $\hat{\mathcal{A}}[\mathbf{z},\l] = \mathcal{A}[(\mathbf{x}(u), -\mathbf{p}(u), \lambda(u))]$, with $u=t-\t$. Since the integration limits coincide under this change of variables, we can relabel the dummy index $du = d\t$, allowing direct subtraction of the integrands \cite{Vaquero2025GSI}. Taking the time derivative gives the medium EPR (see \cite[\S 8.2.1]{Risken1996} for a discussion of time integrals of Markovian variables):
\begin{equation}\label{eq:stoch_s_m}
    \dot s_m = (\dot{p}^\mu - F^\mu_+) D_{\mu\nu} (F^\nu_- -g^\nu_k \partial_{p_\a}g^\a_k)-\partial_{p_\mu}F^\mu_+ ,
\end{equation}
in agreement with \cite{Kwon_2016} when restricting Eq. \eqref{eq:stoch_s_m} to additive noise. Finally, adding both contributions gives the total EPR:
\begin{equation}\label{eq:total_EPR}
\begin{aligned}
    \dot s_{\rm tot} &= \dot s + \dot s_m = -\f{\partial_t \mP}{\mP}-\f{\partial_{x_\mu} \mP}{\mP}\f{p^\mu}{m}+\f{ D_{\mu\nu}j^\nu_d}{\mP}(\dot{p}^\mu-F^\mu_+)\\
    &\phantom{=} - \f{F^\mu_+}{\mP} \partial_{p_\mu}\mP-\partial_{p_\mu}F^\mu_+ .
\end{aligned}
\end{equation}
It can be seen how the stochastic total EPR is not constrained to be nonnegative, since the second law applies on the ensemble level rather than the individual trajectory level.

Computing the average EPR requires a two-step process, involving an average over all trajectories which are at time $t$ in a state $\mathbf{z}$, followed by an ensemble average \cite{Seifert_2005,Seifert_2012}. The first step involves the integral over paths which gives $\avg{\dot{\mathbf{p}}|\mathbf{x},\mathbf{p},t}=\mathbf{j}_p (\mathbf{x},\mathbf{p},t)/\mP(\mathbf{x},\mathbf{p},t)$, required to project the time derivatives defined along a stochastic trajectory into the ensemble space where the PDF $\mP(\mathbf{x},\mathbf{p},t)$ is defined. The second step is taken by averaging over the phase space $\Omega :=\{(\mathbf{x},\mathbf{p})\}$.  As a result, the two-step average yields, for some general function $g$ \cite{Seifert_2012}: 
\begin{equation}
    \avgg{g(\mathbf{x},\mathbf{p}) \mathbf{\dot p}}=\int d\mathbf{x}d\mathbf{p}\, g(\mathbf{x},\mathbf{p})\mathbf{j}_p, 
\end{equation}
where $\mathbf{j}_p$ is the probability current along the momentum dimension defined in \eqref{eq:FP}. See Appendix \ref{app:ensemble_avg} for a proof of this relation. In the following derivations, we take into account the natural boundary conditions associated with the PDF \cite{Risken1996}: 
\begin{equation}\label{eq:natural_BC}
    \mP\big|_{\p \Omega}=\jv \big|_{\p \Omega} = 0.
\end{equation}
The average EPR is then: 
\begin{eqnarray}\label{eq:dot_S_tot}
    \dot S_{\rm tot} (t)= \dot \mS (t) + \dot S_m (t), 
\end{eqnarray}
where we define 
\begin{align}
    &\dot S_{\rm tot} :=\avgg{\dot s_{\rm tot}} = \int d\mathbf{z}\,   \f{j^\mu_d}{\mP} D_{\mu\nu}j^\nu_d, \label{eq:avg_total_EPR} \\
    &\dot S_m := \avgg{\dot s_m} = \int d\mathbf{z}\, j^\mu_d D_{\mu\nu} (F^\nu_- -g^\nu_k \partial_{p_\a}g^\a_k)-\avg{\partial_{p_\mu}F^\mu_+},\label{eq:s_m_definition}
\end{align} 
again consistent with \cite{Kwon_2016} for additive noise. Notably, only the third term in Eq. \eqref{eq:total_EPR} survives the averaging process. Provided the covariance matrix $D^{\nu\mu}$ is positive-definite (PD), the average total EPR yields a quadratic form and hence verifies $\dot S_{\rm tot} \geq 0$, with equality being restricted to equilibrium where dissipative currents vanish $(\jd = \bm{0})$.

Moreover, $\dot \mS = \avgg{\dot s}$, which can be shown by directly differentiating Shannon entropy (see \cite[\S A.3]{strauss2007partial} for total time derivatives of integrals):
\begin{eqnarray}
    \dot \mS  &=& - \int d\mathbf{z}\, \ln \mP\, \f{\p \mP}{\p t} \nonumber\\
&=&  \int d\mathbf{z}\, \ln \mP\,  [\p_{x_\mu}\left(\f{p^\mu}{m}\mP\right) + \p_{p_\mu}  (j^\mu_r + j^\mu_d) ].
\end{eqnarray}
Applying integration by parts, the term $\avg{\p_{\mathbf{x}}\cdot(\mathbf{p}/m)}=0$ vanishes due to independence of phase space variables. The reversible current contribution gives
\begin{equation}
\begin{aligned}
\int d\mathbf{z}\, \ln \mP\,   \p_{\mathbf{p}} \cdot \jr &=\int d\mathbf{z}\, \mP \partial_{p_\mu}\f{j^\mu_r}{\mP} = \avg{\partial_{p_\mu}F^\mu_+}.
\end{aligned}
\end{equation}
For the dissipative current contribution:
\begin{equation}
    \begin{aligned}
    &\int d\mathbf{z}\, \ln \mP\,   \p_{\mathbf{p}}\cdot \jd =  -\int d\mathbf{z}\,    \f{j^\mu_d}{\mP}\p_{p_\mu} \mP \\
    & =  -\int d\mathbf{z}\,    \f{j^\mu_d}{\mP}D_{\mu\nu}\left( (F^\nu_- -g^\nu_k \partial_{p_\a}g^\a_k)\mathcal{P}-j^\nu_d \right)\\
    &= \int d\mathbf{z}\,   \f{j^\mu_d}{\mP} D_{\mu\nu}j^\nu_d   -\int d\mathbf{z}\,   j^\mu_d D_{\mu\nu} (F^\nu_- -g^\nu_k \partial_{p_\a}g^\a_k).
    \end{aligned}
\end{equation}
Gathering the results, the average system EPR gives: 
\begin{equation}\label{eq:dS_multid}
\begin{aligned}
    \dot \mS&=\int d\mathbf{z}\,   \f{j^\mu_d}{\mP} D_{\mu\nu}j^\nu_d   \\
    &\phantom{\dot \mS=}-\int d\mathbf{z}\,   j^\mu_d D_{\mu\nu} (F^\nu_- -g^\nu_k \partial_{p_\a}g^\a_k) +\avg{\partial_{p_\mu}F^\mu_+}\\
     &=\avgg{\dot s_{\rm tot}} - \avgg{\dot s_m} = \avgg{\dot s}.
\end{aligned}
\end{equation}
Hence, while the Shannon entropy is a state function of the PDF, the medium entropy---and consequently, the total entropy---is inherently path-dependent. As such, there exists no state function $S_m$ whose time derivative yields the average medium EP, i.e., $\nexists S_m | \f{d}{dt}S_m = \dot S_m$. The total and medium entropy are therefore defined explicitly through the trajectory-dependent expressions given by Eqs. \eqref{eq:avg_total_EPR} and \eqref{eq:s_m_definition}.

\subsection{Stochastic energetics}\label{sec:ST_energetics}
Stochastic energetics deals with the stochastic extension of the first law and its connections with the second law obtained via ST. We illustrate it with the system Eq. \eqref{eq:LE}, but we distinguish three different kind of forces, $\mathbf{F} = -\nabla_{\mathbf{x}}V(\mathbf{x},\l)-\underline{\underline{\boldsymbol{\gamma}}} (\mathbf{x},\mathbf{p}) \mathbf{p}/m+\mathbf{f}_+(\mathbf{x},\mathbf{p})$, where $V(\mathbf{x},\l)$ gives the potential energy, $\underline{\underline{\boldsymbol{\gamma}}}$ the damping coefficient and $\mathbf{f}_+$ non-conservative even forces, as could be an external manipulation by an agent \cite{Kwon_2016} or feedback cooling \cite{Munakata_2012}. By analogy with the previous section, we identify the reversible and irreversible contributions as $\mathbf{F}_+ = -\nabla_\mathbf{x} V(\mathbf{x},\lambda) +\mathbf{f}_+$ and $\mathbf{F}_- =-\underline{\underline{\boldsymbol{\gamma}}}\cdot\mathbf{p}/m$. Following Sekimoto's approach \cite{Sekimoto2010}, we can derive the first law from the total energy of the system $E$:
\begin{equation}\label{eq:1law}
\begin{aligned}
    dE &= d\left(\frac{|\mathbf{p}|^2}{2m}+V(\mathbf{x},\lambda)\right)=\left(\dot{\mathbf{p}}+\nabla V\right) \cdot\frac{\mathbf{p}}{m}dt+\frac{\partial V}{\partial \lambda}d\lambda \\
    & = ( -\underline{\underline{\boldsymbol{\gamma}}} (\mathbf{x},\mathbf{p})\cdot\mathbf{p}/m +\underline{\underline{\mathbf{g}}}(\mathbf{x},\mathbf{p})\circ\boldsymbol{\zeta}) \cdot \frac{\mathbf{p}}{m}dt\\
    &\phantom{=}  +\mathbf{f}_+ \cdot \frac{\mathbf{p}}{m}dt +\frac{\partial V}{\partial \lambda} d\lambda.
\end{aligned}
\end{equation}
Then, heat and work can be identified as (in units $\kb = 1$) \cite{Seifert_2012,Fodor_2022}:
\begin{subequations}
\begin{eqnarray}
        &&\dot q := ( -\underline{\underline{\boldsymbol{\gamma}}} (\mathbf{x},\mathbf{p})\cdot\mathbf{p}/m +\underline{\underline{\mathbf{g}}}(\mathbf{x},\mathbf{p})\circ\boldsymbol{\zeta})\cdot \frac{\mathbf{p}}{m}, \label{eq:stoch_q}\\
        &&\dot w := \mathbf{f}_+ \cdot \frac{\mathbf{p}}{m} +\frac{\partial V}{\partial \lambda} \dot \lambda. \label{eq:stoch_w}
\end{eqnarray}
\end{subequations}
This identification stems from the fact that heat involves irreversible forces, that is, forces that contribute to entropy production, meanwhile work involves reversible forces that do not explicitly contribute to the total entropy production \cite{Horowitz_2014}. 

In the particular scenario where $\underline{\underline{\boldsymbol{\gamma}}}(\mathbf{x},\mathbf{p})=\underline{\underline{\boldsymbol{\gamma}}}$ is a constant, the Einstein relation $T\underline{\underline{\boldsymbol{\gamma}}}=\underline{\underline{\mathbf{D}}}$, where $\underline{\underline{\mathbf{D}}}$ is the covariance of the random forces and $T$ is the absolute temperature, implies for Eq. \eqref{eq:stoch_s_m}:
\begin{eqnarray}
     \dot s_m &=& -\f{1}{T}( -\underline{\underline{\boldsymbol{\gamma}}} \cdot\mathbf{p}/m +\underline{\underline{\mathbf{g}}}\circ\boldsymbol{\zeta})\cdot \frac{\mathbf{p}}{m} -\nabla_{\mathbf{p}}\cdot \mathbf{f}_+ \nonumber \\
     &=& -\f{\dot q}{T}- \dot \mI,
\end{eqnarray}
recovering the Clausius relation in agreement with the first law Eq. \eqref{eq:1law}, as well as an additional contribution which can be identified as the phase space volume dilation rate difference between the forward and backward trajectories:
\begin{equation}\label{eq:entropy_pump}
    \dot{\mI}=\frac{1}{2}\left[\nabla_{\mathbf{p}}\cdot \mathbf{F}(z)-\nabla_{\hat{\mathbf{p}}}\cdot \mathbf{\hat{F}}(z)\right].
\end{equation}
This term, referred to as entropy pumping, is attributed to the velocity-dependent control force due to an external agent \cite{Kim&Qian}. It is then uncovered to be a lower bound on other information measures, such as information flow, accounting for the information flux due to external manipulation \cite{Horowitz_2014}. In Ref. \cite{Horowitz_2014} the information-theoretic interpretation of entropy pumping was established as a minimal information requirement for the agent to perform the momentum-dependent force $\mathbf{f}_+$, an interpretation that has been further adopted \cite{Kwon_2016,Munakata_2012,Mandal_2017,Baiesi_2015, Chun_2018}. Hence, the EP in the medium includes both dissipative and information terms, as it is well established in the literature \cite{Kim&Qian,Parrondo2015, Dabelow_2019, Cafaro, Sagawa_2012, Esposito_2011, Horowitz_esposito}:
\begin{eqnarray}\label{eq:Gen_clausius}
    \D s_m = -\int\f{dq}{T}-\D \mI ,
\end{eqnarray}
and we refer to Eq. \eqref{eq:Gen_clausius} as the generalized Clausius relation, see Ref. \cite{Horowitz_2014,Kwon_2016}. As a remark, we define the medium as the collection of \textit{all} unobserved DoFs, which include those associated with the system (the heat bath) and those external to it, such as the agent performing work. Accordingly, the heat bath is a subset of the medium.

\subsection{\label{sec:ST_caveats} Thermodynamic consistency}

The average total EP in ST \eqref{eq:KL} has been shown to always satisfy the second law by definition, due to the properties of the KL divergence. The second law can also be derived by applying Jensen's inequality to the IFT, which, however, holds for any Markovian stochastic dynamics that do not necessarily represent a physical process \cite{Seifert_2012}. Thus, we can conclude that this second law is not a true thermodynamical law, since it is not influenced by the thermodynamic consistency of the model. This has been previously pointed out in the literature by identifying the total EP as  an information measure, termed informatic entropy production rate (IEPR), which, although providing a measure of irreversibility, it cannot be associated with thermodynamic quantities such as heat dissipation or the first law \cite{Fodor_2022,Cates_2022,Maes_LDB}. 

On a deeper level, this issue stems from the definition of medium entropy itself in ST. Recall the requirement of LDB is expressed as the condition for all paths $\mathbf{z}(t)$ \cite{Maes_LDB}:
\begin{equation}\label{eq:LDB}
   \ln\frac{P[\mathbf{z}|\mathbf{z}_0 ]}{P[\hat{\mathbf{z}}|\mathbf{z}_t ]}=-\f{\D q}{T} -\D \mI , 
\end{equation}
where the heat bath entropy change is defined as the heat exchanged with the bath divided by its temperature. This notion of entropy has a clear energetic interpretation and coincides with the thermodynamic entropy associated with the bath’s Boltzmann entropy change \cite{xing2025}. This relation holds for any model satisfying LDB \cite{Maes_LDB}, for both continuous- and discrete-time dynamics (see e.g. \cite{Crooks1999Excursions}), and therefore inherits the standard properties of thermodynamic consistency. However, in ST the medium entropy is of this form \textit{by definition} [Eq. \eqref{eq:EP}], and hence it is \textit{blind} to LDB. 

As a consequence, when applying ST to, for instance, active matter or biological systems, terms without a clear thermodynamic interpretation appear in the EP, identified as unconventional EP  \cite{hiddenS,Chun_2018,Kwon_2016,chaudhuri}. This is because such models, of phenomenological origin, do not explicitly account for the driving mechanisms. If not all \textit{thermodynamically} relevant DoFs are taken into account, LDB is violated, the FDRs break down, and entropy production becomes ill-defined due to inconsistent coarse-graining \cite{PhysRevX.15.021050,PhysRevResearch.6.013190,Cates_2022}. Moreover, the lack of thermodynamic consistency can also be found in models that assume the Einstein relation without taking into consideration multiplicative noise or the associated noise-induced drifts, as it was pointed out in Refs. \cite{Lau_2007,Cates_2022}.

Given the requirement of LDB [Eq. \eqref{eq:LDB}], we refer to a thermodynamically consistent model as that which verifies: 
\begin{enumerate}[label=(\alph*)]
    \item Non-decreasing average total EPR $(\dot S_{\rm tot} \geq 0)$ \cite{Eckart,GBYo2019,Esposito_2011}.
    \item The generalized Clausius relation Eq. \eqref{eq:Gen_clausius} such that the thermodynamic entropy has a well-defined physical interpretation \cite{hiddenS,Kwon_2016, Cates_2022, ActiveFields, Horowitz_2014}.
    \item For isolated systems, the FP satisfies an equilibrium steady state \cite{zwanzig,Kardar2007,Mazenko2006,Lau_2007}.
\end{enumerate}
In the following section, we present a variational formulation of (Langevin) ST as a new modeling paradigm which systematically assesses thermodynamic consistency, reconciling the information and physical perspectives of ST.

\section{\label{sec:VP} Variational principle of ST}
\subsection{Thermodynamic phase space}
Following the approach described in Ref. \cite{GBYo2019}, we derive the dynamical equations (ODEs) for both configurational and thermal variables by extending the phase space $\Omega$ to include all thermodynamically relevant DoFs. The fundamental thermal variable introduced in this framework is the stochastic thermodynamic entropy $s$. For example, in a thermomechanical system, the extended phase space becomes $\Omega:=\{(\mathbf{x},\mathbf{p},s)\}$. 

Hence, $s$ is an independent variable that accounts for the (internal) hidden DoFs of the system and is no longer given as a function of $\mP$ as in \eqref{eq:stoch_shannon}. Aligning with the fundamental path-dependence of the medium entropy highlighted in Sec. \ref{sec:stoch_entropy_prod}, $s$ is not assumed to be a function of state \cite{Vaquero2025GSI}; path-independence is recovered near equilibrium, as demonstrated in subsequent sections. This treatment has proven particularly effective in the deterministic (macroscopic) setting for describing nonequilibrium thermodynamics \cite{GBYo2019}, where defining an entropy function is especially challenging \cite{Calazans}. 

In stochastic descriptions---such as Langevin dynamics---the assumption of time-scale separation renders the heat bath concept intrinsic to the formulation. The thermodynamic entropy $s$ thereby characterizes the macrostate of the bath, whose dynamics are governed by heat exchange with the system, $T\,ds = -dq$. This interpretation is consistent with the microscopic dynamical entropy framework introduced in Ref.~\cite{ding2025}, where the bath contribution to the total entropy can be inferred from the system's observable DoFs via the Clausius relation.

In contrast to the standard framework of ST, the PDF is now defined over the complete thermodynamic phase space $\Omega$, acquiring the form $\mP(\mathbf{x},\mathbf{p},s,t)$  following the example of a thermomechanical system. The associated probability measure becomes $\mP(\mathbf{x},\mathbf{p},s,t)d\mathbf{x}d\mathbf{p}ds$. Accordingly, the stochastic system entropy and the Shannon entropy are defined as follows:
\begin{subequations}\label{eq:shannon_VP}
\begin{align}
   & s_{sys}(t)= - \ln \mathcal{P}(\mathbf{x}(t),\mathbf{p}(t),s(t),t),\\
    &\mS(t) = \avg{s_{sys}(t)} = -\int_\Omega d\Omega\, \mP(\mathbf{x},\mathbf{p},s,t)\ln \mP(\mathbf{x},\mathbf{p},s,t),
\end{align}
\end{subequations}
where $d\Omega =d\mathbf{x}d\mathbf{p}ds$. The FP equation is denoted by $\p_t\mP = -\nabla_\Omega\cdot\mathbf{j}$, where $\nabla_\Omega$ denotes the gradient operator with respect to the thermodynamic phase space variables. In this particular example, $\nabla_\Omega=(\p_\mathbf{x},\p_\mathbf{p},\p_s)$. Note that, without assuming an infinite heat bath, the distribution $\mP$ encodes not only the statistics of the observable DoFs but also those of the bath macrostate, through its dependence on the thermodynamic entropy $s$. In subsequent sections it is shown that, under the infinite-bath assumption, the Maxwell–Boltzmann distribution is recovered at equilibrium.

More generally, we denote the thermal variables by $\varphi_\alpha$, which encompass both observable macroscopic quantities---such as the mass $N$ of a chemical species---as well as thermodynamic entropy $s$. These variables span the thermodynamic phase space, which we define as $\Omega := \{(\mathbf{x}, \mathbf{p}, \varphi_\alpha)\} = \{(\boldsymbol{\omega}, s)\}$, where $\boldsymbol{\omega}$ collectively denotes the observable DoFs. Associated with each phase space variable is a conjugate variable, referred to as the (generalized) thermodynamic displacement $\Lambda^\alpha$. These displacements play a central role in our formulation, as they enable the derivation of the general form of the dynamical equations through a variational principle. 

An additional central quantity in this formulation is the medium entropy, denoted by $\Sigma$. This quantity plays a role analogous to the medium entropy $s_m$ in conventional ST, in that it accounts for the entropy production associated with the medium---as defined in Sec. \ref{sec:ST_energetics}. Introducing $\Sigma$ allows us to distinguish unambiguously between entropy changes that contribute directly to the system’s thermodynamic entropy $s$ and those arising from interactions with the external environment. The dynamics associated with $\Sigma$ are determined by the nonlinear nonholonomic constraints, formulated within the framework of the Lagrange-d’Alembert principle. 

\subsection{\label{sec:nonholon} Lagrange-d'Alembert principle}
In Ref. \cite{GBYo2019} the variational approach to nonequilibrium thermodynamics is formulated by constructing a generalization of the Lagrange-d'Alembert principle of
nonholonomic mechanics \cite{Bloch2015}, where the entropy production of the system, written as the sum of the contribution of each of the irreversible processes, is incorporated into a nonlinear nonholonomic constraint. We now extend this formulation to the stochastic regime.

Recall that Hamilton's Principle defines the dynamics of a physical system by the variational principle:
\begin{subequations}  
\begin{eqnarray}
    &&\delta \mathcal{A}[\mathbf{x}] = \delta \int_{t_1}^{t_2}dt L (\mathbf{x}, \mathbf{\dot x})=0,\\
    &&\d \mathbf{x}(t_1) = \d \mathbf{x}(t_2)=0,
\end{eqnarray}
\end{subequations}
where $\mathbf{x}\in V$ stands for generalized coordinates,\footnote{The symbol $V$ denotes the configuration space and, from this point forward, is no longer used to represent the potential energy.} $\mathbf{\dot x} = d\mathbf{x}/dt \in V$ for generalized velocities, and $L(\mathbf{x},\mathbf{\dot x})=K(\mathbf{x},\mathbf{\dot x})-U(\mathbf{x})$ is the Lagrangian function, which contains the physical information in terms of the kinetic energy $K:V\times V\rightarrow \mathbb{R}$ and the potential energy $U:V\rightarrow \mathbb{R}$. Here, we consider the configuration space $V$ as a vector space.

In constructing a Lagrange-d'Alembert principle for ST we first extend the Lagrangian to the complete thermodynamic phase space by including the thermal contributions given by the internal energy $U(\mathbf{x},\varphi_\a)$:
\begin{equation}\label{eq:Lagrangian}
    L(\mathbf{x}, \mathbf{\dot x}, \varphi_\a)=K(\mathbf{x},\mathbf{\dot x})-U(\mathbf{x},\varphi_\a).
\end{equation}
This internal energy encodes the properties of the heat bath, such as its temperature $T(\mathbf{x},\varphi_\a):= \p U/ \p s$ or the exerted force on the system $\mathbf{f}(\mathbf{x},\varphi_\a):= -\p U/\p \mathbf{x}$. These relations correspond the total differential of the internal energy, expressed in terms of the entropy and the observable variables $\bm{\omega}$ as $dU=TdS+ \nabla_{\bm{\omega}}U \cdot d\bm{\omega}$ \cite{chandler1987}. Notably, this definition of temperature arises explicitly from the internal energy and will appear consistently in the FDR. This differs from earlier instances where $T$ appeared merely as a positive parameter enforcing $T\underline{\underline{\boldsymbol{\gamma}}}=\underline{\underline{\mathbf{D}}}$, and was only \textit{a posteriori} interpreted as the bath temperature.

As a technical step, we derive the implicit form of the Euler-Lagrange equations---i.e., the equations for \((\dot{\mathbf{x}}, \dot{\mathbf{p}}, \dot{\varphi}_\alpha)\)---using the Hamilton-Pontryagin principle~\cite{HamiltonPontryagin}. This formulation casts the stochastic dynamics as a first-order system on the thermodynamic phase space, ensuring that all time derivatives are well-defined as stochastic differentials and compatible with the chosen integration convention. In the case of the extended Lagrangian~\eqref{eq:Lagrangian}, the only second-order kinematic constraint is \(\dot{\mathbf{x}} = \mathbf{v}\), with fixed endpoints for \(\mathbf{x}(t)\). Accordingly, we promote \(\mathbf{p}\) to a Lagrange multiplier enforcing this constraint, by augmenting the action with the term \(\int dt\, \mathbf{p} \cdot (\dot{\mathbf{x}} - \mathbf{v})\), consistently with the variational formulation of implicit Lagrangian systems developed in~\cite{HamiltonPontryagin}.

However, to consistently recover the reversible dynamics, the variational principle must yield conservation laws for the thermal variables in the absence of irreversible processes, i.e., $\dot\varphi_\alpha = 0$. This is accomplished by coupling each thermal variable $\varphi_\alpha$ to its conjugate displacement variable $\Lambda^\alpha$ within the action functional, through a term of the form $\int dt\, \varphi_\alpha \dot\Lambda^\alpha$ \cite{GBYo2019}. In the case of entropy, where the time evolution of the medium entropy $\Sigma$ is governed by the nonholonomic constraints, this coupling takes the form $\int dt\, (s-\Sigma) \dot\Gamma$, where $\G$ is the conjugate variable to the entropy $s$. The minus sign in this expression ensures that, in the absence of external fluxes, the variational principle gives $\dot\Sigma = \dot s$,---as expected for a closed system---where the thermodynamic EP accounts for the entire medium entropy. We recall that $s$ corresponds to the entropy production due to heat dissipation, and is distinct from the system entropy $s_{sys}$ associated with the Shannon entropy.

Thus, the complete set of dynamical equations are obtained by taking variations of the (Hamilton-Pontryagin modified) action given by the extended Lagrangian, e.g. Eq. \eqref{eq:Lagrangian},
\begin{align}\label{eq:critical_curve}
    \delta \int_{t_1}^{t_2} dt \bigg( L (\mathbf{x},\mathbf{v},\varphi_\a,s) + \mathbf{p} \cdot ( \mathbf{\dot x} - \mathbf{v})  \nonumber\\
    + \varphi_\a\dot \L^\a+(s-\Sigma)\dot \Gamma \bigg) =0, 
\end{align}
subject to the nonholonomic constraints (to be defined). External manipulations can be accounted for by including a potential $\tilde U(\mathbf{x},\l)$ for a time-dependent driving, or an external non-conservative force $\mathbf{f}_+$ via the principle of virtual work.

To motivate the form of the nonholonomic constraints, we draw an analogy with standard ST, in the same spirit as the geometric formulation of nonequilibrium thermodynamics in Ref.~\cite{GBYo2019}. In our framework, the thermodynamic power associated with an irreversible process mirrors the structure of the heat dissipated by a mechanical force, as expressed in Eq.~\eqref{eq:stoch_q}, or equivalently as $dq = (\mathbf{f}_- + \bm{\xi}) \cdot d\mathbf{x}$. That is, it takes the form of a generalized friction force---comprising both deterministic and stochastic contributions---contracted with a mechanical displacement. 

Translating this structure to the thermodynamic setting, we interpret the thermodynamic fluxes $J_\alpha$ as generalized friction forces, and the differentials of the thermodynamic displacements $d\Lambda^\alpha$ as generalized displacements. This yields an analogous expression for dissipated heat: $dq = J_\alpha\, d\Lambda^\alpha$. By introducing a thermodynamic phase space $\Omega$ equipped with a set of thermal variables, and its dual $\Omega^*$ comprising their corresponding conjugate displacements, this ansatz ensures that the structure of dissipation in general irreversible processes inherits the same geometric form as that of mechanical friction.

With this preamble, we have shown how irreversible processes can be incorporated within the same geometric structure, with entropy production providing the underlying organizing principle. We then embed this structure into a variational formulation by employing the Lagrange-d'Alembert principle \cite{GBYo2019}, in which the critical curve condition is subject to two constraints: a \textit{kinematic} constraint on the solution curve and a \textit{variational} constraint on the variations to be considered when computing the criticality condition. By construction, these constraints encode the information regarding the entropy production. Thus, we define the kinematic and variational nonlinear nonholonomic constraints, respectively, as:
\begin{equation}\label{eq:nonhol_constraints}
\begin{aligned}
    & \f{\p L}{\p s}\dot \Sigma = (J_\a+\xi_\a)\cdot \dot \Lambda^\a + (J_\be+\xi_\be)\cdot (\dot \Lambda^\be-X^\be_{ext}) +  \mathcal{I}_\be\dot\Gamma^\be, \\
    & \f{\p L}{\p s}\d\Sigma=(J_\a+\xi_\a)\cdot \d \Lambda^\a + (J_\be+\xi_\be)\cdot\d \Lambda^\be + \mathcal{I}_\be\delta\Gamma^\be,
\end{aligned}
\end{equation}
where $\a$ and $\be$ denote internal and external processes, respectively, with thermodynamic (irreversible) fluxes $J_\a, J_\be$ and thermodynamic affinities $X^\a, X^\be$, together with a thermodynamic affinity $X^\be_{ext}$ associated with the exterior. Then, the (generalized) thermodynamic displacements $\Lambda^\a,\Lambda^\be$ are such that $\dot \Lambda^\a = X^\a$ and $\dot \Lambda^\be = X^\be$ \cite{GBYo2019}. We recall $\Sigma$ stands for the medium entropy.

In contrast to the deterministic (macroscopic) formulation, each dissipative flux $J_\a,J_\be$ must include its stochastic counterpart $\xi_\a,\xi_\be$ for complete thermodynamic consistency, according to the fluctuation-dissipation theorem (FDT) \cite{zwanzig, Kardar2007, Mazenko2006}. This arises as a consequence of coarse-graining: dissipative forces constitute a macroscopic representation of underlying microscopic processes whose DoFs remain unresolved. The lack of information on these DoFs then has to be explicitly accounted for by the noise in order to correctly quantify the associated entropy production. 

\begin{figure}[h!]
    \centering
    \includegraphics[width=0.7\linewidth]{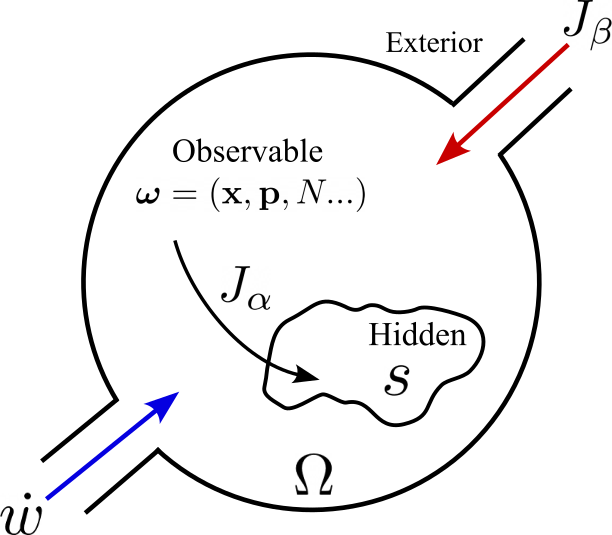}
    \caption{
    The system $\Omega := \{(\bm{\omega}, s)\}$ is closed under internal irreversible processes $J_\alpha$, which involve information loss into the subspace of hidden DoFs represented by the thermodynamic entropy $s$. The system becomes open when subject to external fluxes, exchanging energy and entropy via work $\dot{w}$ and  additional irreversible processes $J_\beta$.
}
    \label{fig:VP_scheme}
\end{figure}

An additional class of thermodynamic fluxes arising in the stochastic framework corresponds to information fluxes. In the context of Langevin dynamics, Sec.~\ref{sec:ST_energetics} showed that external manipulations can induce an entropy flow---termed entropy pumping. As this contribution can be regarded as an external entropy flux $\mI_\be$---see Eq. \eqref{eq:Gen_clausius}---it can be incorporated into the constraint structure \eqref{eq:nonhol_constraints} by coupling it to the conjugate displacement of the thermodynamic entropy $(\G)$, yielding additional terms $\mathcal{I}_p \dot{\Gamma}$ and $\mathcal{I}_p\, \delta\Gamma$ in the kinematic and variational constraints, respectively, where
\begin{equation}\label{eq:entropy_pump_VM}
    \mI_p := \frac{1}{2}\left[\nabla_\Omega\cdot \bm{\mathfrak{F}}-\nabla_{\hat{\Omega}}\cdot \hat{\bm{\mathfrak{F}}}\right],
\end{equation}
and $\bm{\mathfrak{F}}$ collectively denotes all the deterministic forces or fluxes acting on the system (recall $\hat \Box$ denotes time-reversal operation). Equation \eqref{eq:entropy_pump_VM} is just the extension of Eq. \eqref{eq:entropy_pump} to the thermodynamic phase space. Since entropy pumping arises from non-dissipative mechanisms, it is not constrained by the FDT and contributes purely through deterministic terms (i.e., $\xi_{\beta} = 0$). This is consistent with the standard framework of ST, as such contributions do not explicitly contribute to the total EP, see Eq. \eqref{eq:avg_total_EPR}. A concrete example will be provided in Sec.~\ref{sec:mech_open}. See Fig. \ref{fig:VP_scheme} for an illustration of the different thermodynamic fluxes. 

Both type of constraints are systematically related, since the variational constraints can be derived form the kinematic constraints by replacing $d\Lambda^\a \rightsquigarrow\d \Lambda^\a$ and $(d\Lambda^\be-X^\be_{ext}dt) \rightsquigarrow \d \Lambda^\be$, where the exterior contribution vanishes since $\d t = 0$ by definition---variations in Hamilton's principle are taken at fixed time \cite{Bloch2015,GBYo2018}. As a remark, this principle is \textit{not} a critical curve condition for the action integral restricted to the space of a curve satisfying the constraints. However, it is emphasized this formulation recovers Hamilton's Principle in the absence of irreversibility, analogously as in Ref. \cite{GBYo2019}.

Let us summarize and provide more clear insight into the variational structure presented here. On one hand we have defined a thermodynamic phase space on which the system Lagrangian takes values; this allows for a variational treatment of thermal variables, as well as a consistent definition of thermodynamic affinities. The critical curve condition [Eq. \eqref{eq:critical_curve}] has been extended to reproduce conservation laws associated with macroscopic variables. Moreover, we have shown the nonholonomic constraints role is manifold: (i) it allows introducing irreversibility in the equations of motion from the variational point of view; (ii) it does so with a clear geometric structure of contraction between fluxes $J_\g\in \o$ and conjugate displacements $\L^\g\in \o^*$ $[\g=(\a,\be)]$, desirable for a rigorous mathematical setting; (iii) such structure restricts the space of all possible trajectories and variations to the set that have a well-defined energetic interpretation of the associated entropy production. 

Beyond these geometrical aspects, from the physical point of view, the form of the constraints [Eq. \eqref{eq:nonhol_constraints}] define a quantity $\Sigma$ which encodes the ratio of stochastic work done by irreversible forces (and the possible entropy fluxes) and bath temperature, which is just the extended Clausius definition of entropy in ST [Eq. \eqref{eq:Gen_clausius}], as discussed in Sec. \ref{sec:ST_energetics}. Such a definition of $\Sigma$ already implies the mentioned assumptions of time-scale separation, large bath size and micro-reversibility, as required by the stochastic description, the Clausius relation and the well-defined energetics associated with entropy production; see Ref.~\cite{xing2025} for a detailed discussion on the microcanonical ensemble foundations of ST. Crucially, the constraints define $\Sigma$ and not $s$; recall this allows for an unambiguous distinction between thermodynamic entropy production and external entropy flows. This does not imply that the dynamics of $s$ are left unspecified; owing to the critical curve condition [Eq.~\eqref{eq:critical_curve}], the equations of motion for all variables are derived in a self-contained way. 

\subsection{LDB and the second law}\label{sec:LDB_VP}
Up to this point, we have not specified the form of irreversible processes. These are determined by imposing the second law, which automatically determines the FDRs linking the fluxes $\{J_\g\}$ with the noise terms $\{\xi_\g\}$. As it was argued in Sec.~\ref{sec:ST_caveats}, the second law of ST is blind to thermodynamic consistency (or LDB). However, since this framework provides with a well-defined \textit{physical} medium entropy $\Sigma$ consistent with micro-reversibility, the condition of LDB implies reconciling the \textit{physical} definition of $\Sigma$ with the information-theoretic medium entropy $\mathfrak{s}_m$. We denote the (information-theoretic) entropy defined in the extended phase space $\Omega$ as $\mathfrak{s}_m ,\mathfrak{s}_{\rm tot}$ in order to distinguish it from $s_m ,s_{\rm tot}$ in the restricted phase space of Sec. \ref{sec:stoch_entropy_prod}, respectively. That is,
\begin{subequations}\label{eq:inf_entropy_VP}
\begin{align}
    &  \langle \Delta\mathfrak{s}_{\rm tot} \rangle := D_{KL}[P[\mathbf{u}]||P[\hat{\mathbf{u}}]], \\
    &      \Delta\mathfrak{s}_m =\ln\frac{P[\mathbf{u}|\mathbf{u}_0 ]}{P[\hat{\mathbf{u}}|\mathbf{u}_t ]},
\end{align}
\end{subequations}
for a trajectory $\mathbf{u}:[0,t]\subset\mathbb{R}\rightarrow \o$. Recall the system entropy in $\o$ was defined in Eq. \eqref{eq:shannon_VP}.

This can be achieved by defining the second law as a combination of the information $(\mS)$ and physical $(\Sigma)$ entropy contributions, imposed as an axiom on the dynamical equations derived from the variational formulation. Within this framework, the second law is then stated as:
\begin{equation}\label{eq:2nd_law_VP}
    \dot S_{\rm tot}(t) := \dot \mS (t) + \avgg{\dot \Sigma (t)} \geq 0,
\end{equation}
where $\mS(t)$ stands for the Shannon entropy as given by Eq. \eqref{eq:shannon_VP}.  This definition has the same form as that of ST, i.e., Eq. \eqref{eq:dot_S_tot}, except that the inequality is \textit{not} inherently satisfied by definition of $\dot{S}_{\rm tot}$, contrary to the standard definition in ST as given in Eq. \eqref{eq:avg_total_EPR}. Moreover, this expression is defined in the complete thermodynamic phase space $\Omega$. 

To illustrate this result, recall from Sec.~\ref{sec:stoch_entropy_prod} that $\dot \mS (t)$ provides with the information-theoretic contributions [Eq. \eqref{eq:dot_S_tot}], which in the extended phase space $\o$ is denoted as
\begin{equation}\label{eq:Shannon_EPR_VP}
    \avgg{\dot{\mathfrak{s}}_{\rm tot}}=\dot \mS  + \avgg{\dot{\mathfrak{s}}_m}.
\end{equation}
Although $\dot{\mathfrak{s}}_m$ requires path probabilities for its definition and computation at the individual-trajectory level, its ensemble average $\avgg{\dot{\mathfrak{s}}_m}$ can be computed from $\mS$ alone following \eqref{eq:Shannon_EPR_VP}, as shown in \eqref{eq:dS_multid}. 

Then, substituting Eq. \eqref{eq:Shannon_EPR_VP} back in $\dot S_{\rm tot}$:
\begin{equation}\label{all_entropy_rates}
\begin{aligned}
        \dot S_{\rm tot} := \dot \mS  + \avgg{\dot \Sigma }= \avgg{\dot{\mathfrak{s}}_{\rm tot}} - \avgg{\dot{\mathfrak{s}}_m} + \avgg{\dot \Sigma }.
\end{aligned}
\end{equation}
In subsequent examples it will be shown how compliance with the inequality \eqref{eq:2nd_law_VP} requires the solution $\dot{\mathfrak{s}}_m=\dot \Sigma $, or equivalently,
\begin{equation}\label{eq:LDB_VP}
    \D {\mathfrak{s}}_m:=\ln\frac{P[\mathbf{u}|\mathbf{u}_0 ]}{P[\hat{\mathbf{u}}|\mathbf{u}_t ]}=\Delta \Sigma , 
\end{equation}
for a trajectory $\mathbf{u}:[0,t]\subset\mathbb{R}\rightarrow \o$. A formal argument for this result is given in Appendix~\ref{app:positivity}, and subsequent examples demonstrate this explicitly. Establishing a general proof of uniqueness remains an open problem and an interesting direction for future work. Crucially, Eq. \eqref{eq:LDB_VP} determines the FDRs (of the second kind \cite{Maes_LDB}),  which is explicitly shown in Secs. \ref{sec:mech_closed}--\ref{sec:mech_open}.  Moreover, this result implies $  \dot S_{\rm tot}= \avgg{\dot{\mathfrak{s}}_{\rm tot}}$ and hence the information-theoretic and physical descriptions of entropy are reconciled. As a consequence, $s_{\rm tot}$---the individual-trajectory level version of $S_{\rm tot}$ in Eq. \eqref{eq:2nd_law_VP}---inherits the relations satisfied by $\mathfrak{s}_{\rm tot}$, which is further discussed in Sec. \ref{sec:mech_open}.  

Notably,  Eq. \eqref{eq:LDB_VP}---an equality---is the \textit{true} requirement of LDB extended to the thermodynamic phase space $\o$, ensuring its solution provides a consistent physical theory rooted in micro-reversibility. Ultimately, this condition follows from the structure of the nonholonomic constraints \eqref{eq:nonhol_constraints} (i.e., from the definition of $\Sigma$), which encode the fundamental assumptions underlying LDB; see Ref.~\cite{xing2025} for a derivation of LDB. Thus, LDB emerges naturally from the second law axiom [Eq. \eqref{eq:2nd_law_VP}]. We also show through examples how, by construction, this is the only condition required for thermodynamic consistency as defined in Sec. \ref{sec:ST_caveats}, the rest of conditions arising as a consequence of LDB.

Note that the specific nature of noise correlations has not yet been specified. Throughout this work, $\{\xi_\g\}$ are assumed to be zero-mean Gaussian white noises to enable analytical tractability, and our discussion of the results is restricted to this setting. Nevertheless, the Clausius definition of entropy should remain valid for more general stochastic processes \cite{Maes_LDB,PhysRevLett.108.210601}, implying that the structure of the constraints [Eq.~\eqref{eq:nonhol_constraints}] is preserved, comprising a dissipative drift and its associated fluctuations. It remains, however, an open question under which conditions this form may fail or become analytically intractable when enforcing the second law (or characterizing the average EPR), e.g. for arbitrary temporal correlations.

We analyze three different examples: a thermomechanical isolated system (\ref{sec:mech_closed}), an interconnected system of multiple reservoirs (\ref{sec:inter}), and a thermomechanical open system (\ref{sec:mech_open}).

\subsection{\label{sec:mech_closed} Mechanical isolated system}
First, we consider a mechanical isolated system of a particle in a thermal bath, with thermodynamic phase space $\Omega :=\{(\mathbf{x}, \mathbf{p}, s)\}$ and Lagrangian 
\begin{equation}
    L(\mathbf{x}, \mathbf{v}, s)=\frac{1}{2}m|\mathbf{v}|^2-U(\mathbf{x},s)
\end{equation} 
such that $\partial L/\partial s = -\partial U/\partial s = -T(\mathbf{x},s)$. Note the potential $V(\mathbf{x})$ can be absorbed in $U(\mathbf{x},s)$. The variational formulation then involves taking variations of the (Hamilton-Pontryagin modified) action, subject to Lagrange-d’Alembert nonholonomic constraints for the entropy $\Sigma$:
\begin{equation}\label{eq:VP_mech_closed}
    \begin{aligned}
    &\delta \int_{t_1}^{t_2}dt\left( L (\mathbf{x},\mathbf{v},s) + \mathbf{p} \cdot ( \mathbf{\dot x} - \mathbf{v})  + \dot \Gamma(s-\Sigma)\right) =0, \\ 
    &\begin{cases}
   \vspace{0.2cm}\displaystyle\frac{\partial L}{\partial s}\dot \Sigma= (\mathbf{f_-}+ \underline{\underline{\mathbf{g}}}\circ\boldsymbol{\zeta}) \cdot\mathbf{\dot x}, \\
   \vspace{0.2cm}\displaystyle\frac{\partial L}{\partial s}\delta \Sigma= (\mathbf{f_-}+ \underline{\underline{\mathbf{g}}}\circ\boldsymbol{\zeta}) \cdot\delta \mathbf{x}, 
\end{cases}
\end{aligned}
\end{equation}
where $\mathbf{f_-} = -\underline{\underline{\boldsymbol{\gamma}}} (\mathbf{x},\mathbf{p},s) \mathbf{p}/m$ is a dissipative force with state-dependent damping coefficient, $\underline{\underline{\mathbf{g}}}(\mathbf{x},\mathbf{p},s)$ is the noise strength and $\bm{\zeta}(t)$ is gaussian white noise verifying the relations given by Eq. \eqref{eq:gaussian_noise}. $\Gamma$ denotes the (generalized) thermal displacement, i.e. $\dot \Gamma = T$, an equality that follows from the variational principle. 

For simplicity, we assume a 1-dimensional system. Then, the variational principle gives the equations of motion (EoMs):
\begin{equation}\begin{aligned}\label{eq:EoM_mech_closed}
   & \displaystyle p=m\dot{x}, \quad \dot \Gamma = -\f{\p L}{\p s} = T,\\ 
   &  \displaystyle\dot{p}=-\partial_x U(x,s) -\gamma(x,p,s)\f{p}{m}+  g(x,p,s)\circ\zeta,\\
   &  \displaystyle\dot s = \f{-1}{T(x,s)}\left( -\gamma(x,p,s)\f{p}{m}+g(x,p,s)\circ\zeta\right)\f{p}{m},
\end{aligned}\end{equation}
with $\dot s = \dot \Sigma$, equality that holds in any closed system. See Appendix \ref{app:mech_closed} for the detailed derivation. Taking the total time derivative of $E = \frac{1}{2}m|\mathbf{v}|^2 + U(\mathbf{x},s)$ we find the first law 
\begin{equation}\label{eq:1law_mech_closed}
    \dot E = (\dot p + \p_x U)\f{p}{m}+ T\dot s =0,
\end{equation}
 hence energy is conserved and the system is closed. Moreover, the equation for $\dot{s}$ has the same form as Eq. \eqref{eq:stoch_q}, following the Clausius relation and confirming that $s$ consistently represents the thermodynamic entropy, as previously established.

Denoting the dissipative and entropy probability currents, respectively, as
\begin{subequations}
\begin{align}
     j_d =& \left(-\gamma\f{p}{m}-g\partial_p g+\f{p}{m}g\partial_s (g/T)\right)\mP  \nonumber\\
     &-g^2\partial_p \mP+\f{pg^2}{mT}\partial_s \mP, \label{eq:j_d_closed}\\
     j_s =& -\f{p}{mT}\jdd,
\end{align}
\end{subequations}
the system and thermodynamic average EPRs yield (see Appendix \ref{app:mech_closed}):
\begin{align}
    &\dot \mS = \underbrace{\int d\Omega\,\bigg\{ \f{\jdd^2}{\mP g^2}}_{ \avgg{\dot{\mathfrak{s}}_{\rm tot}}} \nonumber\\
     &\phantom{\dot \mS =} - \left. \f{\jdd}{\ g^2} \left(-\gamma\f{p}{m}-g\partial_p g+\f{p}{m}g\partial_s (g/T)\right)\right\} \nonumber\\
    &\phantom{\dot \mS } =  \avgg{\dot{\mathfrak{s}}_{\rm tot}} - \avgg{\dot{\mathfrak{s}}_m} , \label{eq:sys_EPR_closed}\\
    &\avgg{\dot \Sigma}=\avgg{\dot s}=    -\int d\Omega\, \f{p}{mT}\jdd. \label{eq:med_EPR_closed}
\end{align}
The identification of the corresponding terms with $\avgg{\dot{\mathfrak{s}}_{\rm tot}}$ and $\avgg{\dot{\mathfrak{s}}_m}$ follows from a direct verification, see Appendix \ref{app:pseudo_inv}. Adding both contributions gives the total EPR:
\begin{equation}\label{eq:total_EPR_closed}
\begin{aligned}
    \dot S_{\rm tot} &= \dot \mS+\avgg{\dot \Sigma}= \int d\Omega\, \f{\jdd^2}{\mP g^2} \\
    &-\int d\Omega\, \jdd \left\{\f{p}{mT}+\f{1}{g^2}\left(-\gamma\f{p}{m}-g\partial_p g+\f{p}{m}g\partial_s (g/T)\right)\right\}.
\end{aligned}
\end{equation}
The first term is a PD quadratic form, but the second is sign-indefinite; hence, in contrast to standard ST, the total EPR need not be non-negative.

Enforcing the second law requires the vanishing of the term in braces multiplying $j_d$; see Appendix \ref{app:positivity}. From this condition emerges the FDR and the associated total EPR:
\begin{align}
&-\g\f{p}{m}  = -\f{D}{T}\f{p}{m}+g\partial_p g - g\partial_s (g/T)\f{p}{m} \Rightarrow  \label{eq:FDR_mech_closed}\\
 &\dot S_{\rm tot} = \int d\Omega\, \f{\jdd^2}{\mP D}\geq 0 ,   \label{eq:total_EPR_closed_2}
\end{align}
where we recall $D=g^2$. Notably, the total EPR has the same form as that obtained from the standard ST approach in one dimension, see Eq. \eqref{eq:avg_total_EPR}, but it is instead defined over the thermodynamic phase space $\Omega$. The corresponding multidimensional version of Eq. \eqref{eq:FDR_mech_closed}, obtained by an analogous computation, is
\begin{equation}\label{eq:FDR_mech_closed_multi}
    -\g^{\nu\mu}\f{p_\mu}{m}  = -\f{D^{\nu\mu}}{T}\f{p_\mu}{m}+g^\nu_k \partial_{p_\mu}g^\mu_k - g^\nu_k\partial_s (g^\mu_k/T)\f{p_\mu}{m}.
\end{equation}

As anticipated in Sec.~\ref{sec:LDB_VP}, this solution, or equivalently, the associated FDR, enforces the identity $\dot \Sigma = \dot{\mathfrak{s}}_m$, see Eqs.~\eqref{eq:sys_EPR_closed} and \eqref{eq:total_EPR_closed_2}. Hence, LDB [Eq.~\eqref{eq:LDB_VP}] emerges directly from the second law axiom of the variational principle, without recourse to path probabilities. For completeness, Appendix~\ref{app:pseudo_inv} reproduces the same result via an explicit path probability calculation.

Remarkably, since the thermodynamic EP satisfies
\begin{equation}
\begin{aligned}
    \f{d}{dt}\avg{s} &= \int d\Omega\, s \partial_t \mP = -\int d\Omega\, s \nabla_\Omega \cdot\mathbf{j} = \int d\Omega\, \mathbf{j}\cdot\nabla_\Omega s \\
    &= \int d\Omega\, j_s=-\int d\Omega\, \f{p}{mT}\jdd =\avgg{\dot s},
\end{aligned}
\end{equation}
it implies that  $\dot S_{\rm tot}$ is a total time derivative, as we can write
\begin{equation}\label{real_time_rate}
   \dot S_{\rm tot}= \frac{d}{dt}(\mathcal{S}+  \langle s\rangle),
\end{equation}
in contrast with the standard ST setting, i.e. the results of Sec. \ref{sec:ST}. Thus, for isolated systems, the entropy behaves as a function of state, as it is expected near equilibrium.

The FDR [Eq. \eqref{eq:FDR_mech_closed}] connects the dissipative forces with the random forces covariance, plus contributions from noise-induced drift terms. In the special case of constant temperature and constant damping coefficient, we recover the standard Einstein relation discussed in Sec. \ref{sec:ST_energetics}, namely $T\gamma = D$. 

When $\gamma,T,g$ are independent of the entropy $s$, our result reproduces the FDR derived in Ref. \cite{Dubkov2009}, where this relation was derived by setting the steady-state distribution to be the Boltzmann distribution. This reference also illustrates how momentum-dependent multiplicative noise can arise, for instance, in systems with Coulomb friction, which requires an extension of the FDR to ensure thermodynamic consistency. Additionally, in this scenario---or more generally, for $\p_p {\rm f}_+=0$---where the observable DoFs decouple from the entropy balance in \eqref{eq:EoM_mech_closed}, $\dot s$ recovers the same expression as $\dot s_m$ in Eq. \eqref{eq:stoch_s_m} under the FDR constraint \eqref{eq:FDR_mech_closed_multi}, providing it with a well-defined thermodynamic interpretation (Clausius relation).

Beyond recovering these known results, this framework introduces a novel approach for coupling mechanical and thermal DoFs \cite{Vaquero2025GSI}. This allows for taking into account finite baths and/or coupled system-bath dynamics. The third contribution in Eq. \eqref{eq:FDR_mech_closed}, the noise-induced drift due to $s$ dependence, is essential to guarantee thermodynamic consistency in the presence of such nonlinear system-bath coupling.

Addressing the existence of an equilibrium steady state distribution, in order to complete the requirements of thermodynamic consistency provided in Sec \ref{sec:ST_caveats}, by imposing the FDR \eqref{eq:FDR_mech_closed} we find that the functional 
\begin{equation}\label{eq:SS_functional}
    \mP(x,p,s)=f\left(E(x,p,s)\right)e^s
\end{equation}
satisfies the FP at steady state for any function of the total energy $f(E)$. Computing the current components gives:
\begin{subequations}  
\begin{align}
    &\partial_x j_x = \f{p}{m}\partial_x \mP = \f{p}{m}e^s f'(E)\partial_x U ,\\
    & \partial_p j_p = \partial_p \left(-\partial_x U\mP-\f{pg^2}{mT}\mP+\f{pg^2}{mT}\mP\right)= -\partial_x j_x,\\
     &\partial_s j_s = \partial_s \left(\left[\f{pg}{mT}\right]^2\mP-\left[\f{pg}{mT}\right]^2\mP\right)=0,\\
     &\nabla_\Omega \cdot\mathbf{j} = \partial_x j_x - \partial_x j_x = 0,
\end{align}
\end{subequations}
which implies $\partial_t \mP=-\nabla_\Omega\cdot\mathbf{j}=0$. The distribution \eqref{eq:SS_functional} is general and holds for coupled system-bath dynamics as well. Moreover, it favors the states that maximize the entropy $s$, consistent with the maximum entropy principle in statistical mechanics, as well as being of the same form as the equilibrium distribution in fluctuation theory \cite{chandler1987,Ruppeiner,Landau,Einstein}, with an additional prefactor accounting for energy conservation. 

An explicit form of Eq. \eqref{eq:SS_functional} can be derived by using the fact that the energy distribution $\mP(e)$ is time-independent, given energy is conserved, see Eq. \eqref{eq:1law_mech_closed}:
\begin{equation}
\begin{aligned}
        \mP_0(e)&=\mP(e)=\int_{\Omega}d\Omega\, f(E(x,p,s))e^s\d (e-E(x,p,s))  \\
        &=f(e)\int_\Omega d\Omega\, e^s\d (e-E(x,p,s)) = f(e)Z(e) \\
        & \Rightarrow f(e) = \f{ \mP_0(e)}{Z(e)} .
\end{aligned}
\end{equation}
Due to the time-independence of $f(e)$, taking logarithms on both sides of Eq. \eqref{eq:SS_functional} and taking the difference at $t_1, t_2$ gives 
\begin{equation}
    \D s = \D \ln \mP= - \D s_{sys} \Rightarrow \D s_{\rm tot} = \D s_{sys}+\D s = 0,
\end{equation}
where $s_{sys}$ is given by Eq. \eqref{eq:shannon_VP} and $s_{\rm tot}$ is the individual-trajectory level version of $S_{\rm tot}$. Thus the stochastic total entropy production vanishes along any trajectory at equilibrium.

The phase space integral for $Z(e)$ can be solved applying the property \cite{Arfken}:
\begin{equation}\label{eq:composite_Dirac_delta}
    \d (g(x))=\sum_i \f{\d (x-x_i)}{|g'(x_i)|} \quad \text{for}\quad g(x_i)=0,
\end{equation}
for which an explicit form of $E(x,p,s)$ is required. In the simplest scenario where $T=\p_s E$ is a parameter, we assume $U(x,s)=Ts$, such that the particle and bath energies are separable $E = H_0 (x,p) + Ts$. Denoting $s_0=(e-H_0(x,p))/T$:
\begin{equation}
    \begin{aligned}
        Z(e)&=\int_\Omega d\Omega\, e^s\d (e-E(x,p,s)) =\int_\Omega dxdpds \, e^s\f{\d (s-s_0)}{|\partial_s E|}\\
        &=\int_{X,P} dxdp \,\f{1}{|T|}e^{\f{1}{T}(e-H_0(x,p))}\\
        & = \f{1}{|T|}e^{e/T}\int_{X,P} dxdp \,e^{-H_0(x,p)/T} =\f{1}{|T|}e^{e/T}\; Z_{MB}.
    \end{aligned}
\end{equation}
Hence,
\begin{eqnarray}\label{eq:MB}
    \mP (x,p,s) &=& \mP_0(e)\f{e^s}{Z(e)}\d (e-E(x,p,s)) \nonumber\\
    &\propto&  \mP_0(E(x,p,s))\f{e^{s-e/T}}{Z_{MB}} \nonumber \\
    &=& \mP_0(E(x,p,s))\f{e^{-H_0(x,p)/T}}{Z_{MB}}.
\end{eqnarray}
This corresponds precisely to the Maxwell-Boltzmann (MB) canonical distribution \cite{chandler1987,Landau}, subject to energy conservation. We note that, for a uniform $\mP_0(e)$, the distribution becomes independent of $s$. This situation corresponds to an infinite bath whose macroscopic state remains unchanged. 

Remarkably, in this formulation, the MB distribution emerges as a consequence of the FDR derived from the second law of thermodynamics, contrary to the conventional approach, where the FDR is obtained by assuming the MB distribution as the steady-state. Hence, the variational approach recovers the results of equilibrium statistical mechanics in the canonical ensemble, without being restricted to such regime---recall we do not require the assumption of infinite baths for the FDRs nor the equilibrium ansatz [Eq. \eqref{eq:SS_functional}].

\subsection{\label{sec:inter} Interconnected system of multiple reservoirs}
\begin{figure}[h!]
    \centering
    \includegraphics[width=0.8\linewidth]{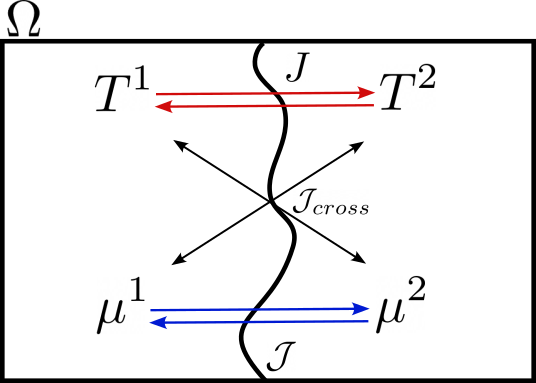}
    \caption{Schematic of an interconnected system with two reservoirs exchanging entropy ($J$) and mass ($\mJ$). Cross-effects induced by cross-correlated noise lead to additional fluxes $\mJ_{cross}$, enabling heat (mass) transfer in response to chemical potential (temperature) gradients.}
    \label{fig:inter}
\end{figure}

In this example, we consider multiple reservoirs interacting via stochastic fluxes. For simplicity, let us assume two reservoirs that exchange mass and heat, such that the whole system is closed. The entropy and mass variables for each reservoir are denoted $\{(s_i ,N_i)\}$ for $(i=1,2)$, respectively. The Lagrangian corresponds to $L=-U(s_1,s_2,N_1,N_2)$, which gives the temperature and chemical potential of each reservoir by
\begin{align}
    T^k (s_1,s_2,N_1,N_2)= \f{\partial U}{\partial s_k} , \\
    \mu^k (s_1,s_2,N_1,N_2)=  \f{\partial U}{\partial N_k}
\end{align}
accordingly. The mass and entropy fluxes are given by $\mJ_{12}=-\mJ_{21}$ and $J_{12}=J_{21}$ respectively, with the stochastic counterparts given by the noises $\zeta(t),\eta(t)$ with amplitudes $\s=\s_{12}=-\s_{21}$ and $\kappa=\kappa_{12}=\kappa_{21}$. These symmetries are derived from total mass and energy conservation due to the system being closed. For the purpose of the variational formulation, it is convenient to define the flux $J_{kl}$ for $k=l$ as \cite{GBYo2018} 
\begin{equation}\label{modeling_assumption}
    J_{kk}:= -\sum_{l\neq k}J_{kl},
\end{equation}
so that $\sum_k J_{kl}=0$. The associated constraints are then given by
\begin{equation}
    \frac{\partial L}{\partial s_k}\d \Sigma_k=\sum_l(J_{kl}+\k_{kl}\circ\eta)\d\Gamma^l +\sum_l (\mJ_{l k}+\s_{l k}\circ\zeta)\d W^k .
\end{equation}
Thus, the variational formulation takes the form
\begin{multline}
    \delta \int_{t_1}^{t_2} dt \left(L(s_1,s_2,N_1,N_2) +  \dot{W}^1 N_1 + \dot{W}^2 N_2 \right.\\
   \hfill \left. +\dot\Gamma^1(s_1-\Sigma_1)+\dot\Gamma^2(s_2-\Sigma_2) \right)=0,\\
   \begin{cases}
   \vspace{0.2cm}\displaystyle\frac{\partial L}{\partial s_1}\dot \Sigma_1=(J_{12}+\kappa_{12}\circ\eta)(\dot\Gamma^2-\dot\Gamma^1) \\
   \hfill -(\mJ_{12}+\s_{12}\circ\zeta)\dot W^1 ,\\
\vspace{0.2cm}\displaystyle \frac{\partial L}{\partial s_2}\dot \Sigma_2=(J_{12}+\kappa_{12}\circ\eta)(\dot\Gamma^1-\dot\Gamma^2) \\
\hfill +(\mJ_{12}+\s_{12}\circ\zeta)\dot W^2,\\
\vspace{0.2cm}\displaystyle\frac{\partial L}{\partial s_1}\d \Sigma_1=(J_{12}+\kappa_{12}\circ\eta)(\d\Gamma^2-\d\Gamma^1)\\
\hfill -(\mJ_{12}+\s_{12}\circ\zeta)\d W^1, \\
\vspace{0.2cm}\displaystyle \frac{\partial L}{\partial s_2}\d \Sigma_2=(J_{12}+\kappa_{12}\circ\eta)(\d\Gamma^1-\d\Gamma^2)\\
\hfill +(\mJ_{12}+\s_{12}\circ\zeta)\d W^2,
\end{cases} 
\end{multline}
where the variables $\Gamma^i,W^i$ are conjugate variables with $s_i$ (or $\Sigma_i$) and $N_i$, respectively. Taking variations then gives the following system (see Appendix \ref{app:inter}):
\begin{equation}\label{eq:EoM_interc}
    \begin{aligned}
       & \dot s_k = \dot \Sigma_k ,\\
       & \dot N_1 = -\mJ_{12}-\s_{12}\circ\zeta ,\\
       & \dot N_2 = \mJ_{12}+\s_{12}\circ\zeta,\\
        & T^1\dot s_1=(J_{12}+\kappa_{12}\circ\eta)(T^1-T^2)+(\mJ_{12}+\s_{12}\circ\zeta)\mu^1 ,\\
       &  T^2 \dot s_2=(J_{12}+\kappa_{12}\circ\eta)(T^2-T^1)-(\mJ_{12}+\s_{12}\circ\zeta)\mu^2,
    \end{aligned}
\end{equation}
where fluxes and noises are allowed to be state-dependent, i.e. $\mJ_{12}=\mJ_{12}(s_1,s_2,N_1,N_2),\,\s_{12}=\s_{12}(s_1,s_2,N_1,N_2)$, and similarly for $J_{12},\kappa_{12}$. Moreover, for the sake of generality, we further allow the noises to be cross-correlated. Denoting $\bm{\xi}=(\zeta,\eta)$:
\begin{equation}
    \avg{\bm{\xi}(t)\otimes\bm{\xi}(t')} = 2 \begin{bmatrix}
        1 & C\\
        C & 1
    \end{bmatrix}\d (t-t'),
\end{equation}
where $C$ denotes the cross-correlation coefficient. It will be shown how cross-correlation implies a common hidden physical mechanism behind the heat and mass fluxes, giving rise to cross-effects (see Fig. \ref{fig:inter}). The same reasoning applies to coupled-particle systems (see, e.g.,~\cite{10.21468/SciPostPhys.17.4.096,PhysRevLett.116.068301}), where analogous cross-effects arise naturally.

The first law in this system gives, applying Eqs. \eqref{eq:EoM_interc}:
\begin{equation}
    \dot E = \mu^k \dot N_k + T^k \dot s_k = 0,
\end{equation}
such that energy is conserved and the system is closed.

To analyze the EPR we define the following currents, according to the FP equation $(\p_t\mP = -\nabla_\Omega\cdot\bm{j})$ for cross-correlated gaussian white noise \cite{cross_FP}:
\begin{subequations}\label{eq:currents_inter}
\begin{align}
    &j_{N_1} = -\tilde{\mJ}_{12}\mP - \s^2 (\p_{N_1}\mP-\p_{N_2}\mP) \nonumber\\
    &\phantom{j_{N_1} =} -\s^2 \left(\f{\mu^2}{T^2}\p_{s_2}\mP-\f{\mu^1}{T^1}\p_{s_1}\mP \right) \nonumber\\ 
    &\phantom{j_{N_1} =} +C\s\kappa\left(\f{T^1-T^2}{T^1}\p_{s_1}\mP+\f{T^2-T^1}{T^2}\p_{s_2}\mP\right) ,\\
    &j_{N_2}  = -j_{N_1},\\
    &j_{s_1} = -\frac{\mu^1}{T^1}j_{N_1}+\f{T^1-T^2}{T^1}j_{th},\\
    &j_{s_2} = -\frac{\mu^2}{T^2}j_{N_2}+\f{T^2-T^1}{T^2}j_{th}.
\end{align}
\end{subequations}
For clarity, we have denoted the thermal contribution as 
\begin{equation}\begin{aligned}\label{eq:j_th_inter}
    &j_{th} = \tilde{J}_{12}\mP-\kappa^2(1-C^2)\left(\f{T^1-T^2}{T^1}\partial_{s_1} \mP+\f{T^2-T^1}{T^2}\p_{s_2}\mP\right)\\
    & \phantom{j_{th} = } -\f{C\kappa}{\s}\left(\tilde{\mJ}_{12}\mP + j_{N_1}\right),
\end{aligned}\end{equation}
and absorbed the noise-induced drifts in
\begin{subequations}   
\begin{align}
    \tilde{\mJ}_{12} = \mJ_{12}+ \nu_m, \\
    \tilde{J}_{12} = J_{12} + \nu_{th},
\end{align}
\end{subequations}
where the explicit expressions for $\nu_m ,\nu_{th}$ can be found in Appendix \ref{app:inter}. The average system and thermodynamic EPRs then yield, respectively,
\begin{widetext}
\begin{align}
    &\dot \mS  = \int \f{d\Omega}{\mP} \begin{bmatrix}
        j_{N_2} \\ j_{th}
    \end{bmatrix}^T  \mathbb{L}^{-1}\left\{ \begin{bmatrix}
        j_{N_2} \\ j_{th}
    \end{bmatrix}- \begin{bmatrix}
        \tilde{\mJ}_{12}\mP \\ \tilde{J}_{12}\mP
    \end{bmatrix}\right\} =  \avgg{\dot{\mathfrak{s}}_{\rm tot}} - \avgg{\dot{\mathfrak{s}}_m}, \label{eq:S_sys_inter}\\
    & \avgg{\dot \Sigma}=\sum_k\avgg{\dot s_k} = \int d\Omega\left\{j_{N_2}\left(\frac{\mu^1}{T^1}-\frac{\mu^2}{T^2}\right)+j_{th}\left(\f{T^1-T^2}{T^1}+\f{T^2-T^1}{T^2}\right)  \right\},\label{eq:med_EPR_inter}
\end{align}
\end{widetext}
with \begin{equation}\label{eq:L_inter}
    \mathbb{L} = \begin{bmatrix}
       \vspace{2mm} \s^2 & C\kappa\s \\ C\kappa\s & \kappa^2
    \end{bmatrix},\; \mathbb{L}^{-1} = \f{1}{(1-C^2)}\begin{bmatrix}
        \vspace{1mm}\displaystyle\f{1}{\s^2} & \displaystyle-\f{C}{\kappa\s} \\ \displaystyle-\f{C}{\kappa\s} & \displaystyle\f{1}{\kappa^2}
    \end{bmatrix},
\end{equation}
being symmetric PD matrices. The identification of the corresponding terms with $\avgg{\dot{\mathfrak{s}}_{\rm tot}}$ and $\avgg{\dot{\mathfrak{s}}_m}$ follows from a direct verification, see Appendix \ref{app:inter}. Note that Eq.  \eqref{real_time_rate} is again satisfied since $\sum_k\avgg{\dot s_k}= \frac{d}{dt} \sum_k\avg{s_k}$, and thus $S_{\rm tot}$ behaves as a state function.

Adding both contributions, the average total EPR gives the form, with the shorthand notation $\bm{\mathfrak{J}}=(j_{N_2},j_{th})$:
\begin{align}\label{eq:total_EPR_inter}
    \dot S_{\rm tot} &= \dot\mS + \avgg{\dot \Sigma}= \int \f{d\Omega}{\mP}\bm{\mathfrak{J}}^T\mathbb{L}^{-1}\bm{\mathfrak{J}} \nonumber\\
    & - \int d\Omega\,\bm{\mathfrak{J}}^T\left\{ \mathbb{L}^{-1}
    \begin{bmatrix}
        \tilde{\mJ}_{12} \\ \tilde{J}_{12}
    \end{bmatrix}-
    \begin{bmatrix}
        \vspace{2mm}\displaystyle\frac{\mu^1}{T^1}-\frac{\mu^2}{T^2} \\ \displaystyle\f{T^1-T^2}{T^1}+\f{T^2-T^1}{T^2}
    \end{bmatrix}\right\}.
\end{align}

Following the general argument in Appendix \ref{app:positivity}, enforcing the second law requires that the term in braces vanish, yielding the FDRs and associated total EPR:
\begin{align}
& \begin{bmatrix}
     \vspace{1.5mm}\tilde{\mJ}_{12} \\
     \tilde{J}_{12}
\end{bmatrix} = \mathbb{L} \begin{bmatrix}
   \vspace{2.5mm}\displaystyle \frac{\mu^1}{T^1} - \frac{\mu^2}{T^2} \\
   \displaystyle (T^1-T^2)\left(\f{1}{T^1}-\f{1}{T^2}\right)
\end{bmatrix}\Rightarrow \label{eq:FDR_inter} \\
& \dot S_{\rm tot}=\int \f{d\Omega}{\mP}\bm{\mathfrak{J}}^T\mathbb{L}^{-1}\bm{\mathfrak{J}}\geq 0 . \label{eq:S_tot_inter}
\end{align}
The average total EPR is a PD quadratic form, attaining zero only at equilibrium where all dissipative currents vanish. Moreover, the FDRs enforce $\dot \Sigma = \dot{\mathfrak{s}}_m$, see Appendix~\ref{app:inter} for an explicit path probability calculation. Hence, fully analogous to the mechanical example discussed in Sec. \ref{sec:mech_closed}, LDB [Eq. \eqref{eq:LDB_VP}] emerges directly from the second law axiom.

 The matrix $\mathbb{L}$ has eigenvalues
\begin{equation}
    \l_{\pm}=\f{1}{2}\left(\kappa^2+\s^2\pm\sqrt{(\s^2-\kappa^2)^2+4C^2\kappa^2\s^2}\right),
\end{equation}
which satisfy $\l_{\pm}>0$ for $|C|<1$, ensuring that $\mathbb{L}$ is PD. In this case, its inverse $\mathbb{L}^{-1}$ is also PD, since its eigenvalues are $1/\l_{\pm}>0$, which guarantees that $\dot S_{\rm tot} \geq 0$. The special case $C=\pm 1$ leads to a degenerate situation in which the two noise terms become identical, introducing a redundancy and rendering the model ill-defined.

The FDR \eqref{eq:FDR_inter} shows that, due to cross-correlations, each flux is influenced by both temperature and chemical potential gradients, thereby exhibiting cross-effects. The relationship between the fluxes and the gradients is governed by the symmetric PD matrix $\mathbb{L}$, which embodies the Onsager reciprocal relations. Indeed, the condition of LDB implies micro-reversibility, from which Onsager symmetry emerges \cite{Onsager_relations,Broeck_1,Broeck_2,Polettini}. This structure indicates that, in the macroscopic limit, the model inherently recovers the phenomenological relations of nonequilibrium thermodynamics \cite{Onsager_relations,GBYo2018}, mediated by the Onsager matrix. Namely, the FDR \eqref{eq:FDR_inter} has the same form as its macroscopic counterpart found in \cite{GBYo2018} from
\begin{equation}\label{I_macroscopic}
    I:= \mJ_{kl}\left(\frac{\mu^k}{T^k} - \frac{\mu^l}{T^l}\right) + J_{kl}(T^l-T^k)\left(\f{1}{T^l}-\f{1}{T^k}\right)\geq 0,
\end{equation}
where $I$ denotes the macroscopic internal EP. In addition, the fluxes include noise-induced drifts, arising from state-dependent correlations, which are necessary for thermodynamic consistency. These features are particularly relevant in modeling active field theories, as discussed in Refs. \cite{Cates_2022, ActiveFields}.  
Since the system is closed, the functional 
\begin{equation}\label{eq:SS_inter}
    \mP(s_1,s_2,N_1,N_2)=f(E(s_1,s_2,N_1,N_2))e^{\sum_k s_k}
\end{equation} 
satisfies the FP at steady state, analogously as in Sec. \ref{sec:mech_closed}, see Eq. \eqref{eq:SS_functional}. This can be shown by enforcing the FDRs \eqref{eq:FDR_inter} in the currents given by Eqs. \eqref{eq:currents_inter}, together with the relations
\begin{subequations}
\begin{eqnarray}
    &&\p_{N_i}\mP = e^{\sum_k s_k}f'(E)\mu^i,\\
    &&\p_{s_i}\mP = e^{\sum_k s_k}f'(E)T^i + \mP.
\end{eqnarray}
\end{subequations}
Substituting in the probability currents yields:
\begin{equation}
    j_{N_i}=0,\ j_{th}=0 \Rightarrow \p_t \mP = -\nabla_\Omega \cdot \bm{j} = 0.
\end{equation}
Again, the system displays an equilibrium steady state under the FDRs derived through the second law, as we require for the thermodynamic consistency of isolated systems. The functional \eqref{eq:SS_inter} can be further simplified for specific forms of the energy $E(s_1,s_2,N_1,N_2)$ by applying the composite rule for Dirac delta functions given by Eq. \eqref{eq:composite_Dirac_delta}. 

For example, assuming that the reservoirs have equilibrated at a temperature $T=T^1=T^2$ and the Helmholtz free energy $F(N_1,N_2)$ is separable, the energy function takes the form $E(s_1,s_2,N_1,N_2)=F(N_1,N_2) + T (s_1 + s_2)$, where $\p_{N_i}F=\mu^i(N_1,N_2)$. The phase space integral $Z(e)$ then gives:
\begin{equation}
    \begin{aligned}
        Z(e)&=\int_\Omega d\Omega\, e^{s_1 + s_2}\d (e-E(s_1,s_2,N_1,N_2))  \\
        &\propto e^{e/T}\int_{N_1,N_2} dN_1dN_2 \,e^{-F(N_1,N_2) /T}=e^{e/T}\; Z_{N}.
    \end{aligned}
\end{equation}
The explicit form of the equilibrium distribution yields
\begin{align}\label{ED_exchanger}
    \mP (s_1,s_2,N_1,N_2) &=\mP_0(E(s_1,s_2,N_1,N_2))\f{e^{s_1+s_2}}{Z(E)} \nonumber\\
    &\propto   \mP_0(E(s_1,s_2,N_1,N_2))\f{e^{-F(N_1,N_2)/T}}{Z_N}.
\end{align}
Analogously to the example in Sec. \ref{sec:mech_closed}, this distribution favors states that minimize the total chemical free energy, subject to energy conservation. Assuming a uniform initial energy distribution, the prefactor can be absorbed into the normalization constant, giving the finite-dimensional analog distribution of Refs.  \cite{Derrida,Leonard_2013}.

\subsection{\label{sec:mech_open} Mechanical open system}
In this section, we consider a particle driven by an external agent through a force $\mathbf{f}_+ (\mathbf{x},\mathbf{p},\l)$, immersed in a heat bath undergoing external heat transfer; see, for instance, Refs.~\cite{PhysRevLett.89.050601,Li2019,DasPRE} for relevant experimental and numerical realizations. A key distinction from the previous examples is that the system is not constrained to relax to a steady state and may remain far from equilibrium throughout its evolution. The corresponding Lagrangian is given by 
\begin{equation}
    L(\mathbf{x}, \mathbf{v}, s)=\frac{1}{2}m|\mathbf{v}|^2-U(\mathbf{x},s).
\end{equation}
The variational formulation proceeds by taking variations of the (Hamilton–Pontryagin modified) action, subject to the nonholonomic Lagrange–d’Alembert constraints \eqref{eq:nonhol_constraints} and incorporating the principle of virtual work to account for the external manipulation: 
\begin{equation}
    \begin{aligned}
    &\delta \int_{t_1}^{t_2} dt \left( L (\mathbf{x},\mathbf{v},s) + \mathbf{p} \cdot ( \mathbf{\dot x} - \mathbf{v})  + \dot \Gamma(s-\Sigma) \right)  \\
     &\hspace{5cm}  +\int_{t_1}^{t_2} dt\; \mathbf{f}_+\cdot \delta \mathbf{x}=0, \\ 
    &\begin{cases}
   \vspace{0.2cm}\displaystyle\frac{\partial L}{\partial s}\dot \Sigma= (\mathbf{f_-}+ \underline{\underline{\mathbf{g}}}\circ\boldsymbol{\zeta}) \cdot\mathbf{\dot x}+(\mathcal{J}_{s,h}+\k\circ\eta)\cdot(\dot \Gamma-T^h)\\
   \hfill +\mI_p\, \dot\Gamma, \\
   \vspace{0.2cm}\displaystyle\frac{\partial L}{\partial s}\delta \Sigma= (\mathbf{f_-}+ \underline{\underline{\mathbf{g}}}\circ\boldsymbol{\zeta}) \cdot\delta \mathbf{x}+(\mathcal{J}_{s,h}+\k\circ\eta)\cdot\delta\Gamma \\
   \hfill +\mI_p\, \d\Gamma.
\end{cases}
\end{aligned}
\end{equation}  
The dissipative mechanical forces are interpreted as in Sec. \ref{sec:mech_closed}. We denote the thermal displacement by $\Gamma$, and the corresponding external entropy flux by $\mJ_{s,h}(\mathbf{x}, \mathbf{p}, s)$, with its stochastic counterpart given by $\k(\mathbf{x}, \mathbf{p}, s) \circ \eta(t)$. The external thermodynamic affinity at the boundary is represented by $T^h$, the temperature of the external reservoir. This should not be confused with the case of an open thermodynamic system with \textit{prescribed fluxes}. Additionally, we incorporate the entropy pumping due to external manipulation, denoted by $\mI_p$, as defined in Eq. \eqref{eq:entropy_pump_VM}.

\begin{figure}[h!]
    \centering
    \includegraphics[width=0.8\linewidth]{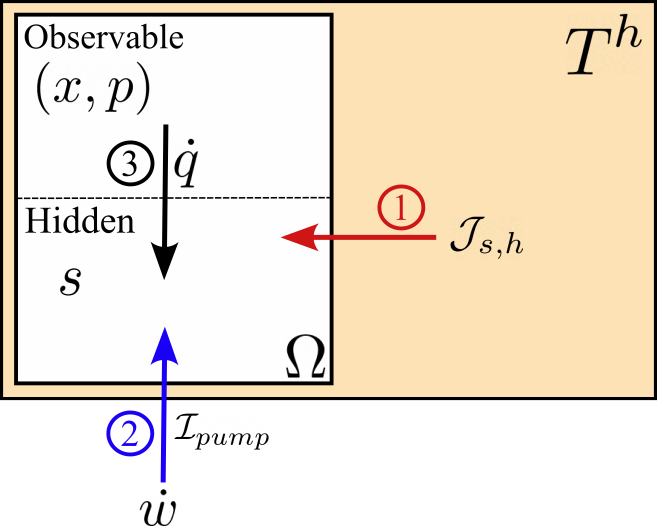}
    \caption{Schematic of a thermomechanical open system. (1) denotes external heat flux, (2) represents entropy pumping (associated with work exchange), and (3) indicates internal dissipation. The entropy flow $\dot{s}_f$ captures contributions external to the system $\Omega$, corresponding to mechanisms (1) and (2). The thermodynamic entropy $s$ accounts for heat dissipated via processes (1) and (3). In contrast, the medium EPR $\dot{\Sigma}$ includes all contributions from (1), (2), and (3). Together, $\dot{\Sigma}$ and the system EPR $\dot \mS$ characterize the total entropy production in the super-system comprising $\Omega$, the thermal reservoir $T^h$, and the external agent.}
    \label{fig:open}
\end{figure}

We restrict to 1-dimension for convenience, giving the EoMs (see Appendix \ref{app:mech_open}):
\begin{equation}\begin{aligned}\label{eq:EoM_open}
    &p=m\dot{x},\quad \dot \Gamma = -\f{\p L}{\p s} = T,\\ 
   & \displaystyle\dot{p}=-\partial_x U-\gamma(x,p,s)\f{p}{m}+{\rm f}_+ +  g(x,p,s)\circ\zeta,\\
   & \displaystyle\dot s =\f{-1}{T(x,s)}\left( -\gamma(x,p,s)\f{p}{m}+g(x,p,s)\circ\zeta\right)\cdot \f{p}{m} \\
    &\phantom{\dot s =} \displaystyle +\f{T^h}{T}\cdot(\mathcal{J}_{s,h}+\k\circ\eta), \\
    & \displaystyle\dot \Sigma =\f{-1}{T(x,s)}\left( -\gamma(x,p,s)\f{p}{m}+g(x,p,s)\circ\zeta\right)\cdot \f{p}{m} \\
    & \phantom{\dot \Sigma =} \displaystyle +\f{T^h - T}{T}\cdot(\mathcal{J}_{s,h}+\k\circ\eta)-\mI_p,\\
     &\dot s_f =\dot s-\dot \Sigma = \mathcal{J}_{s,h}+\k\circ\eta+\mI_p ,
\end{aligned}\end{equation}
where $T$ represents the internal system temperature, to be distinguished from the external bath temperature $T^h$. Since the system is open, the medium EPR $\dot{\Sigma}$ includes an additional contribution to the thermodynamic entropy, accounting for entropy exchanges with sources outside the system $\Omega$. We denote this contribution by $\dot{s}_f$, representing the entropy flow entering or leaving the system. See Fig. \ref{fig:open} for an illustration of the system described by Eqs. \eqref{eq:EoM_open}.

Regarding the first law we find
\begin{align}\label{eq:1law_open}
    \dot E &=\left(\dot p+\p_x U\right)\cdot \f{p}{m}+\p_s U\dot s={\rm f}_+\cdot \f{p}{m}+(\mathcal{J}_{s,h}+\k\circ\eta)T^h \nonumber\\ 
    &=P^{ext}_W + P^{ext}_{H}.
\end{align}
Here, $P^{ext}_W,P^{ext}_{H}$ denote the external power input due to mechanical work and heat transfer, respectively. Equation \eqref{eq:1law_open} represents the stochastic extension of its deterministic counterpart, as given in Eq. (22) of \cite{GBYo2018}.

We denote the ensemble currents associated with the mechanical and thermal irreversible fluxes as
\begin{subequations}\label{eq:currents_open}
\begin{align}
     &\jdd = \left(-\gamma\f{p}{m}-g\partial_p g+\f{p}{m}g\partial_s (g/T)\right)\mP \nonumber\\
     & \phantom{\jdd =} -g^2\partial_p \mP+\f{pg^2}{mT}\partial_s \mP, \\
    & j_{th} = \left[\mathcal{J}_{s,h}-\k\p_s\left(\f{T^h}{T}\k\right)\right]\mP-\f{T^h}{T}\k^2\f{\partial \mP}{\partial s},
\end{align}
\end{subequations}
which define the FP probability currents $j_p,j_s$ shown in Appendix \ref{app:mech_open}. The physical intuition behind Eqs. \eqref{eq:currents_open} is illustrated via the averaged first law:
\begin{equation}\label{avg1law}
\begin{aligned}
    \f{d}{dt}\avg{E}& = -\int d\o\; E\; \nabla_\o\cdot\bm{j} = \int d\o\;  \bm{j}\cdot \nabla_\o E\\
    & = \int d\o\;  \left({\rm f}_+ \cdot \dfrac{p}{m} \mP_t+j_d\cdot \dfrac{p}{m} + Tj_s\right)\\
    &=\avg{{\rm f}_+ \cdot \dfrac{p}{m}}+\int d\o\;   T^h j_{th}.
\end{aligned}
\end{equation}
It can be seen how Eq. \eqref{avg1law} is consistent with the average of Eq. \eqref{eq:1law_open}, and $j_{th}$ corresponds to the ensemble description of the physical entropy flux $(\mathcal{J}_{s,h}+\k\circ\eta)$.
 
The average system EPR gives (see Appendix \ref{app:mech_open})
\begin{equation}
\begin{aligned}
    \dot \mS =& \underbrace{\int \f{d\Omega}{\mP}\,\left(\f{\jdd^2}{g^2}+\f{j_{th}^2}{\k^2}\right)}_{\avgg{\dot{\mathfrak{s}}_{\rm tot}}} +\avg{\partial_p {\rm f}_+ } \\
    & -\int d\Omega\, \f{\jdd}{\ g^2} \left(-\gamma\f{p}{m}-g\partial_p g+\f{p}{m}g\partial_s (g/T)\right)\\
    & -\int d\Omega\,\f{j_{th}}{\k^2}\left[\mathcal{J}_{s,h}-\k\p_s\left(\f{T^h}{T}\k\right)\right]\\
    =&  \avgg{\dot{\mathfrak{s}}_{\rm tot}} - \avgg{\dot{\mathfrak{s}}_m}.
\end{aligned}
\end{equation}
The identification of the corresponding terms with $\avgg{\dot{\mathfrak{s}}_{\rm tot}}$ and $\avgg{\dot{\mathfrak{s}}_m}$ follows from a direct verification, see Eq. \eqref{eq:PI_sm_open} below. Given that $\mI_p = \p_p {\rm f}_+$, the average medium EPR yields
\begin{equation}\label{eq:med_EPR_open}
\begin{aligned}
     \avgg{\dot \Sigma} &=\avgg{\dot s}-\avgg{\dot s_f} \\
    &= \int d\Omega\, \left(-\f{p}{mT}\jdd+ \f{T^h-T}{T}j_{th}\right)-\avg{\p_p {\rm f}_+}. 
\end{aligned}
\end{equation}
Adding both contributions the total average EPR gives
\begin{equation}\label{eq:total_EPR_open}
    \begin{aligned}
        \dot S_{\rm tot} &= \dot\mS + \avgg{\dot \Sigma}\\
        &= \int \f{d\Omega}{\mP}\,\left(\f{\jdd^2}{g^2}+\f{j_{th}^2}{\k^2}\right)\\
        & -\int d\Omega\, \jdd \left\{\f{p}{mT}+\f{1}{g^2}\left(-\gamma\f{p}{m}-g\partial_p g+\f{p}{m}g\partial_s (g/T)\right)\right\}\\
    & +\int d\Omega\,j_{th}\left\{\left(\f{T^h-T}{T}\right)-\f{1}{\k^2}\left[\mathcal{J}_{s,h}-\k\p_s\left(\f{T^h}{T}\k\right)\right]\right\}.
    \end{aligned}
\end{equation}

Following the general argument in Appendix \ref{app:positivity}, enforcing the second law requires that the terms in braces should vanish, giving the FDRs and associated total EPR:
\begin{align}
    &\begin{cases}
        \vspace{3mm}\displaystyle -\gamma\f{p}{m}=-\f{D}{T}\f{p}{m}+g\partial_p g-\f{p}{m}g\partial_s (g/T),\\
        \vspace{3mm}\displaystyle \mathcal{J}_{s,h}=\k^2\left(\f{T^h -T}{T}\right)+\k\p_s\left(\f{T^h}{T}\k\right),
    \end{cases} \label{eq:FDR_open} \\
    &\Rightarrow \dot S_{\rm tot} =  \int \f{d\Omega}{\mP}\,\left(\f{\jdd^2}{D}+\f{j_{th}^2}{\k^2}\right) \geq 0 .
\end{align}
The FDR for the mechanical DoFs coincides with that of a closed mechanical system discussed in Sec. \ref{sec:mech_closed}, see Eq. \eqref{eq:FDR_mech_closed}. This is because, in the present example, we have not included cross-effects between the dissipative force and the external heat flux, although such couplings could, in principle, be incorporated. Additionally, the FDR associated with the external heat flux resembles that of interconnected systems introduced in Sec. \ref{sec:inter}, see Eq. \eqref{eq:FDR_inter}, corresponding to the discrete Fourier's law. In the macroscopic limit, it again converges to the results of nonequilibrium thermodynamics, i.e., the FDR \eqref{eq:FDR_open} is consistent with its macroscopic counterpart given by \cite{GBYo2018}
\begin{equation}
    I:=  -\f{1}{T}\left(-\g\f{p}{m}\right)\cdot \f{p}{m}+\mJ_{s,h}\left(\f{T^h -T}{T}\right)\geq 0.
\end{equation}

In Sec. \ref{sec:LDB_VP} we anticipated that the FDRs obtained through the second law axiom restore LDB by ensuring the identity $\dot{\mathfrak{s}}_m = \dot \Sigma$, and we further demonstrated this result throughout Secs. \ref{sec:mech_closed}-\ref{sec:inter}. We now confirm this result in the far from equilibrium setting by providing an explicit demonstration and examining its implications. 

Applying the PI formulation to the system described by Eqs. \eqref{eq:EoM_open}, we compute the ratio of probabilities for forward and backward trajectories in the thermodynamic phase space $\Omega$. The inverse of the covariance matrix is of the form
\begin{align}
D_{\mu\nu}=\begin{bmatrix}
    \vspace{2mm}\displaystyle\f{1}{g^2}+\f{(p/m)^2}{\k^2(T^h)^2} & \displaystyle\f{T(p/m)}{\k^2(T^h)^2}\\
     \displaystyle\f{T(p/m)}{\k^2(T^h)^2} &  \displaystyle\f{T^2}{\k^2(T^h)^2}
\end{bmatrix}.
\end{align}
Let us define the even and odd vectors 
\begin{subequations}
\begin{align}
    \bf u_+ &= \begin{bmatrix}
    \dot p +\partial_x U - {\rm f}_+\\
     \displaystyle\f{p}{mT}(-\gamma\f{p}{m}-\nu_p)- \f{T^h}{T}( \mathcal{J}_{s,h}-\k\p_s(\k T^h /T))
\end{bmatrix}, \\
    \bf u_- &= \begin{bmatrix}
    \displaystyle\gamma p/m+\nu_p \\ \dot s
\end{bmatrix},
\end{align}
\end{subequations}
with the noise-induced drift $\nu_p = g\partial_p g-\f{p}{m}g\partial_s (g/T)$. We denote $F^p =-\partial_x U-\gamma p/m+{\rm f}_+,\ F^s = \f{\g}{T}(p/m)^2 +  \f{T^h}{T} \mathcal{J}_{s,h} $ as the deterministic forces in the momentum and entropy Eqs. \eqref{eq:EoM_open}. Decomposing $D_{\mu\nu}$ into the symmetric and anti-symmetric contributions $D_{\mu\nu}^+ ,D_{\mu\nu}^-$, respectively, the OM action and its time reversal are given by 
\begin{subequations}
\begin{align}
     \mathcal{A}[\mathbf{u}] =& \int_{t_1}^{t_2}d\tau\left( \f{1}{4}(\zrr+\zii)^T D_{\mu\nu}(\zrr+\zii) \right.\nonumber\\
        & \left. +\f{1}{2}(\p_p F^p + \p_s F^s)\right),\\
      \hat{\mathcal{A}}[\mathbf{u}] =& \int_{t_1}^{t_2}d\tau\left( \f{1}{4}(\zrr-\zii)^T (D_{\mu\nu}^+ - D_{\mu\nu}^-)(\zrr-\zii)\right. \nonumber\\
      & \left. +\f{1}{2}(-\p_p \hat F^p + \p_s \hat F^s)\right),
\end{align}
\end{subequations}
where $\mathbf{u}=(x,p,s)$. The log ratio \eqref{eq:EP} then gives
\begin{align}
     \hat{\mathcal{A}}[\mathbf{u}]- \mathcal{A}[\mathbf{u}] =& \int_{t_1}^{t_2}d\tau \left(\f{1}{4}\left( -2\zrr D_{\mu\nu}^-\zrr-2\zii D_{\mu\nu}^-\zii \right.\right. \nonumber\\
      &-\left.4\zrr D_{\mu\nu}^+ \zii\right) -\p_pF_+^p - \p_s F_-^s \bigg ).
\end{align}
Thus, after some algebra,
\begin{align}
    \D \mathfrak{s}_m &= \int_{t_1}^{t_2}d\tau\left( \f{1 }{g^2}\left( -\gamma\f{p}{m}+g\circ\zeta\right)\left(-\gamma\f{p}{m}-\nu_p \right)\right. \nonumber\\
    & \left. +  \f{1}{\k^2}(\mathcal{J}_{s,h}+\k\circ\eta)( \mathcal{J}_{s,h}-\k\p_s(\k T^h /T)) - \p_pF_+^p \right).
\end{align}
Differentiating and enforcing the FDRs \eqref{eq:FDR_open}, we find
\begin{align}\label{eq:PI_sm_open}
    \dot{\mathfrak{s}}_m  =& \f{d}{d\tau}\text{ln}\frac{P[\mathbf{u}|\mathbf{u}_0 ]}{P[\hat{\mathbf{u}}|\mathbf{u}_t ]} \nonumber \\
     =& -\f{1}{T}\left( -\gamma\f{p}{m}+g\circ\zeta\right)\cdot \f{p}{m}
    +  \f{T^h -T}{T}(\mathcal{J}_{s,h}+\k\circ\eta) \nonumber\\
    &- \p_p {\rm f}_+ ,
\end{align}
which comparing with Eqs. \eqref{eq:EoM_open} gives the identity $\dot{\mathfrak{s}}_m = \dot \Sigma$. Crucially, the variational formulation distinguishes the medium EPR $\dot\Sigma$ from the thermodynamic EPR $\dot s$ by properly accounting for external entropy flows. As a result, it correctly identifies $\dot{\mathfrak{s}}_m = \dot \Sigma$, rather than $\dot{\mathfrak{s}}_m = \dot s$.

This verification establishes the key results anticipated in Sec.~\ref{sec:LDB_VP} in the far from equilibrium regime:
(i) it confirms the interpretation of $\dot{\Sigma}$ as the medium EPR, and of the FDRs as the link between the second law axiom and LDB [Eq. \eqref{eq:LDB_VP}];
(ii) it shows that the FDRs provide a well-defined physical meaning to $\dot{\mathfrak{s}}_m$ by ensuring the generalized Clausius relation \eqref{eq:Gen_clausius};
(iii) it demonstrates that, under the FDRs, the decomposition $\D s_{\rm tot} = \D s_{sys}+\D \Sigma$, with
\begin{equation}
    \D s_{sys}=\text{ln}\frac{\mathcal{P}(\mathbf{u}_0, t_0)}{\mathcal{P}(\mathbf{u}_t,t)} ,\quad \Delta \Sigma  =\text{ln}\frac{P[\mathbf{u}|\mathbf{u}_0 ]}{P[\hat{\mathbf{u}}|\mathbf{u}_t ]},
\end{equation}
is consistent with the KL divergence representation of entropy production \eqref{eq:inf_entropy_VP}, i.e. $\langle \Delta s_{\rm tot}\rangle = \langle \Delta\mathfrak{s}_{\rm tot}\rangle $. 

As a consequence, we can prove the following relation, for arbitrary (normalized) initial and final distributions $\mP_0 ,\mP_t$ \cite{Seifert_2012}:
\begin{align}\label{eq:masterFT}
    \avg{e^{-\ln \mP_0/\mP_t - \D \Sigma}} &= \avg{e^{- \D \Sigma} \mP_t /\mP_0 } \nonumber \\
    &= \int \mathcal{D}\mathbf{u}\,\mP_0 P[\mathbf{u}|\mathbf{u}_0]e^{- \D \Sigma} \mP_t /\mP_0  \nonumber \\
    & = \int \mathcal{D}\mathbf{u}\, P[\hat{\mathbf{u}}|\mathbf{u}_t ]\mP_t = 1,
\end{align}
where the last equality follows due to normalization, analogously as for the IFT, see \eqref{eq:IFT} in Sec. \ref{sec:stoch_entropy_prod}. The relation \eqref{eq:masterFT}, referred to as the master fluctuation theorem (master FT), serves as a unifying result from which many known FTs in ST can be derived as special cases; see Ref. \cite{Seifert_2012} for a detailed discussion. This demonstrates that a significant portion of the ST literature is encompassed within the variational framework presented here, rather than constituting an independent framework. 

Notably, the resulting FTs remain valid even in the presence of nonlinear couplings between thermal and mechanical DoFs, thereby broadening the domain of applicability of current formulations. Moreover, the meaning of the master FT is elevated: in our framework the quantity appearing in the exponent, $\Delta \Sigma$, acquires a clear physical interpretation as the medium entropy change, rather than being understood solely as an information-theoretic term. Ultimately, Eq.~\eqref{eq:masterFT} is a direct consequence of LDB [Eq.~\eqref{eq:LDB_VP}] in the extended phase space, and thus applies to all examples considered in this work. 

A technical remark is in order: when applying the same computation to systems lacking external irreversible processes, such as in the example of Sec. \ref{sec:mech_closed}, one often encounters a singular covariance matrix due to degeneracies arising from energy conservation. In these cases, the master FT can still be recovered by employing the Moore–Penrose pseudo-inverse \cite{meyer01}, as illustrated in Appendix \ref{app:pseudo_inv}.

\section{\label{sec:discussion} Discussion}

Through the examples presented above, we have demonstrated how the variational formulation of ST provides a systematic route to thermodynamically consistent modeling. In particular, enforcing the second law axiom within the variational principle ensures consistency in the precise sense defined in Sec.~\ref{sec:ST_caveats}, namely, compatibility with LDB. Imposing the non-negativity of the average total EPR directly yields the corresponding FDRs, or equivalently, the LDB condition Eq.~\eqref{eq:LDB_VP}, and thus, the thermodynamic entropy is well defined through the Clausius relation and, for isolated systems, the associated FP equation admits an equilibrium steady state. Crucially, none of these properties is imposed \textit{ad hoc}: they arise naturally from the structure of the variational construction itself.

Although Appendix~\ref{app:positivity} provides a formal argument for the emergence of FDRs as solutions of the second law inequality within the variational framework, and their consistency is verified throughout Secs.~\ref{sec:mech_closed}--\ref{sec:mech_open}, a rigorous proof of uniqueness remains an open problem. This observation points to a promising direction for future work: elucidating the general correspondence between the second law axiom and LDB, and clarifying its relation to alternative constraints, such as the orthogonality condition arising in generalized Langevin equations~\cite{Jung2021}. Moreover, although derived from a geometrically motivated ansatz, the dynamical equations governing entropy production---including those at the level of individual trajectories---retain the same form as in conventional ST, up to a reformulation in an extended phase space. Remarkably, once the FDRs are specified, this formulation coincides with the KL–divergence approach, thereby ensuring consistency with fluctuation theorems and compatibility with the broader theoretical framework.

In Secs. \ref{sec:mech_closed} and \ref{sec:inter}, it is shown that the equilibrium distribution arises as a special case of a general equilibrium functional \eqref{eq:SS_functional}, which recovers classical results---namely, the MB distribution---in appropriate limits. A natural direction for future work is to explore equilibrium distributions in more general settings beyond the MB case and to compare the resulting predictions with experimental observations, for instance in coupled system–bath dynamics.

The derived FDRs generalize the standard formulations by allowing a new class of couplings between thermal and configurational DoFs. In the examples discussed in Secs.~\ref{sec:mech_closed}–\ref{sec:mech_open}, these couplings represent system–bath interactions, whereas in Sec.~\ref{sec:inter} they correspond to bath–bath interactions. Within this framework, thermodynamic affinities such as temperature and chemical potential remain well defined, and the bath dynamics themselves follow from the same variational principle. Hence, the underlying \textit{mathematical} structure remains consistent even in regimes where LDB and the definition of thermodynamic affinities are typically difficult to guarantee \cite{Bogers_2025,Maes_FDR}. 

This is due to the fact that the thermodynamic phase space accounts for the complete super-system in a top-down fashion, analogous to the microcanonical formulation \cite{xing2025}, in contrast to conventional ST formulations where the bath is not explicitly included, leading to ambiguities and requiring heat and entropy to be inferred indirectly from the system dynamics. In the bottom-up approach of Ref.~\cite{xing2025}, developed under the same fundamental assumptions as here, it is shown that explicitly accounting for the full super-system allows ST to be extended to strongly-coupled systems and tunable system–bath interactions. Crucially, the LDB condition remains unchanged, and the resulting theory retains the same structural form as in the weak-coupling regime \cite{PhysRevE.109.034105}, suggesting that the present approach should be tested in such regimes as well. Another shared feature is that LDB can be derived without recourse to path probabilities.

Nonetheless, the physical meaning and range of applicability of this extended description remain open questions, indicating an interesting direction for future research involving both experimental and theoretical validation. For instance, in the free energy variational framework of Ref.~\cite{free_energy_VM}, it is shown that a Legendre transform of the action shifts this entropy dependence into an explicit temperature dependence of the parameters, thereby allowing systematic inclusion of cases where phenomenological coefficients---such as viscosity---vary with temperature. Such couplings become particularly relevant in systems where this temperature dependence is non-negligible, as demonstrated, for example, in Ref.~\cite{Diamant_2015}.

We emphasize that, due to the intrinsic dependence of the probability distribution on thermodynamic entropy, steady states of the Fokker-Planck equation necessarily correspond to equilibrium states, as they imply a stationary (time-independent) entropy. Nonetheless, this does not contradict previous findings on nonequilibrium steady states; the key difference is that those steady states refer to marginal distributions where the entropy state variable has been integrated out.

A further advantage of this approach is its capacity to model complex systems as interconnections of primitive subsystems. As demonstrated in Sec. \ref{sec:inter}, this modular construction naturally yields FDRs for coupled mass and heat transfer that define symmetric, PD matrices consistent with Onsager symmetry \cite{Onsager_relations}. In the presence of cross-correlated noise, these matrices consistently capture cross-effects in accordance with nonequilibrium thermodynamic theory and LDB.

The variational formulation also extends to open systems operating far from equilibrium. In the macroscopic limit, the obtained FDRs are consistent with phenomenological results from classical nonequilibrium thermodynamics \cite{GBYo2018,GBYo2019}. Importantly, this formalism allows model parameters---such as temperature, transport coefficients or friction---to be state-dependent in the complete phase space $\Omega$, thus offering significant flexibility. 

A number of technical subtleties arise in specific limits. For example, while we have focused on the underdamped regime to retain the full dynamical structure, taking the overdamped limit within the variational framework must be done with care. In particular, the operations of enforcing the second law and applying the overdamped approximation do not, in general, commute: the overdamped approximation should be performed prior to the second law constraint. However, even then, the overdamped approximation does not provide a generally reliable account of entropy production and may fail to capture key contributions \cite{entropy_anomaly}.  

Furthermore, in the presence of multiplicative noise, the noise-induced drift terms lead to FDRs that take the form of partial differential equations rather than algebraic constraints. While closed-form solutions may be obtainable in specific cases---as in Ref. \cite{Dubkov2009} for Coulomb friction---this is not guaranteed in general. In such cases, the model can be systematically extended by incorporating these drifts into an effective dissipative force (e.g., ${\rm f}_- \rightarrow {\rm f}_- + \nu_p$), resulting in an algebraic FDR. This strategy has been employed in various contexts \cite{Lau_2007,Cates_2022,ActiveFields}, and fits naturally within the variational formulation developed here. 

Finally, although this framework requires an initial specification of the thermal variables, any inconsistency between this choice and the underlying dynamics will arise as a mathematical inconsistency. For instance, in systems exhibiting negative friction, the FDR implies $g^2/T = \gamma < 0$, which is incompatible with $g^2 \ge 0$ unless one assumes a negative temperature $T < 0$. However, since temperature is defined independently of $\gamma$, such a condition cannot be imposed. Consequently, LDB breaks down, appearing as an inconsistency in the FDRs. This apparent violation can be resolved by recognizing that an effective energy input arises from coupling to an additional DoF---meaning that the initial model neglected a thermodynamically relevant variable~\cite{PhysRevX.15.021050}. Hence, the framework does not arbitrarily enforce LDB but instead reveals when a more complete thermodynamic phase space must be considered, thereby guiding the modeling process.

\section{\label{sec:conclusion} Conclusion}
The variational formulation of ST provides a unified and systematic geometric framework for constructing models that are thermodynamically consistent by design. Crucially, this approach remains valid far from equilibrium, applies to both closed and open systems, and naturally accommodates nonlinear bath interactions through coupling of configurational and thermal variables. It systematically integrates microscopic fluctuations with macroscopic thermodynamic structure, offering a flexible and modular formalism grounded in the second law. These features make it broadly applicable, from isolated mechanical systems to open, interacting, and actively driven environments.

Remarkably, FDRs have emerged as fundamental constraints for thermodynamic consistency to ensure compliance with the second law of thermodynamics, providing a well-defined interpretation of entropy production and ensuring LDB. Within this context, the variational approach offers a systematic framework to derive such relations from first principles, under the assumptions of time-scale separation, large bath size and micro-reversibility. Although we have restricted our results to Gaussian noise, we expect the nonholonomic constraints to be independent of the particular noise correlations. Exploring this direction---by extending the formalism to incorporate non-Markovian or non-Gaussian dynamics, for instance through generalized FP equations~\cite{Denisov_2009}---remains an interesting avenue for future research. Finally, as a variational formulation rooted in Hamilton’s principle, the current framework naturally applies to Langevin-type dynamics, but not to more general Markov chains.

Ongoing work extends this framework to infinite-dimensional systems (fields), where the underlying geometric structure of the variational principle enables thermodynamically consistent modeling of complex fluids, active matter, and other continuum systems. Promising research directions include the development of variational integrators for structure-preserving numerical schemes \cite{Gawlik2020,Gawlik_2021,gawlik24}, the application of Lagrangian reduction by symmetry to capture emergent phenomena \cite{GAYBALMAZ2009,reduction_nematic}, a
systematic treatment of conservation laws and Lie group symmetries, and, as previously mentioned, the role of configurational-thermal variables coupling and non-Markovian dynamics.

\begin{acknowledgments}
H. V. del P. acknowledges the Provost Graduate Award scholarship from Nanyang Technological University, Singapore, and thanks J. M. R. Parrondo for valuable discussions on information thermodynamics.  F. G.-B. is partially supported by a startup grant from Nanyang Technological University and by the Ministry of Education, Singapore, under Academic Research Fund (AcRF) Tier 1 Grant RG99/24. H.Y. is partially supported by JSPS Grant-in-Aid for Scientific Research (22K03443), JST CREST (JPMJCR24Q5), and Waseda University (SR 2025C-095). The authors thank the Referees for their careful review and constructive suggestions, which have improved the manuscript.

H. V. del P. derived the results and prepared the original draft. F. G.-B. and L. Y. C. supervised the work. F. G.-B., H. Y., and L. Y. C. contributed to the review and editing of the manuscript. All authors participated in the discussions.
\end{acknowledgments}

% \clearpage
\appendix
\section{\label{app:ensemble_avg} Two-step average proof}
In order to prove
\begin{equation}
    \avgg{g(\mathbf{x},\mathbf{p}) \mathbf{\dot p}}=\int d\mathbf{x}d\mathbf{p}\, g(\mathbf{x},\mathbf{p})\mathbf{j}_p, 
\end{equation}
we only need to prove $\avg{\dot{\mathbf{p}}|\mathbf{x},\mathbf{p},t}=\mathbf{j}_p (\mathbf{x},\mathbf{p},t)/\mP(\mathbf{x},\mathbf{p},t)$, since
\begin{equation}
\begin{aligned}
    \avgg{g(\mathbf{x},\mathbf{p}) \mathbf{\dot p}}&=\avg{g(\mathbf{x},\mathbf{p}) \avg{\dot{\mathbf{p}}|\mathbf{x},\mathbf{p},t}}\\
    &= \int d\mathbf{x}d\mathbf{p}\,\mP(\mathbf{x},\mathbf{p},t)g(\mathbf{x},\mathbf{p})\f{\mathbf{j}_p (\mathbf{x},\mathbf{p},t)}{\mP(\mathbf{x},\mathbf{p},t)}\\
    &= \int d\mathbf{x}d\mathbf{p}\,g(\mathbf{x},\mathbf{p})\mathbf{j}_p .
\end{aligned}
\end{equation}
We also restrict to the 1-dimensional case for clarity. Following the approach in \cite{Seifert_2012}, we take the 2nd order central difference consistent with Stratonovich convention:
\begin{eqnarray}\label{eq:center_difference}
    \avg{\dot{p}|x,p,t}=&&\lim_{\epsilon\rightarrow 0}\f{1}{2\epsilon}\left[\avg{p(t+\epsilon)-p(t)|x,p,t}\right. \nonumber\\
    && + \left.\avg{p(t)-p(t-\epsilon)|x,p,t}\right],
\end{eqnarray}
where the average over paths is performed over one single time step $\D t=\epsilon$ through the discretized propagator \eqref{eq:PI_multid}, conditioned on $(x(t)=x,p(t)=p)$. We assume $\dot p$ is taken at some fixed position, such that we can treat $x$ as a constant. The change of variables to $\D p =p(t+\epsilon)-p(t)$ simplifies the conditional probability through the PI integral to \begin{equation}
    \begin{aligned}
        & P(p(t+\epsilon)=p+\D p\,|\,p(t)=p) \\
        & =\frac{1}{\sqrt{4 \pi g^2 \epsilon}}\exp\left[ -\left\{\frac{1}{4 g^2\epsilon}\left(\D p-(f-gg')\epsilon\right)^2+\frac{\epsilon}{2}f' \right\} \right]\\
        &=  \frac{1}{\sqrt{4 \pi g^2 \epsilon}}\exp\left[ -\left\{\left(\alpha-\beta\right)^2+\frac{\epsilon}{2}f' \right\} \right],
    \end{aligned}
\end{equation}
with $\a=\D p/\sqrt{4 g^2 \epsilon}$ and $\beta=(f-gg')\epsilon/\sqrt{4 g^2 \epsilon}$.
Hence, the first term of Eq. \eqref{eq:center_difference} yields:
\begin{equation}
    \begin{aligned}
        &\avg{p(t+\epsilon)-p(t)\,|\,x,p,t}= \int_\mathbb{R} d(\D p)\D p\, P(p+\D p|p)\\
        & = \exp\left(-\frac{\epsilon}{2}f'\right)\sqrt{\frac{4g^2\epsilon}{\pi }}\int_\mathbb{R}d\a  \,\a e^{ -\left(\alpha-\beta\right)^2}\\
        &=\exp\left(-\frac{\epsilon}{2}f'\right)\sqrt{4 g^2 \epsilon}\beta= \exp\left(-\frac{\epsilon}{2}f'\right)(f-gg')\epsilon.
    \end{aligned}
\end{equation}
where the properties of the gaussian integral have been applied \cite{Risken1996}. 

For the second term of Eq. \eqref{eq:center_difference}, the endpoint conditioning $p(t)=p$ is applied using Bayes theorem \cite{Seifert_2012,Arfken}. Since the associated Langevin equation is a Markov process, we apply Bayes theorem in the context of Markov chains as \cite{Kampen}:
\begin{equation}
    \begin{aligned}
        &P(p(t-\epsilon)=p-\D p\,|\,p(t)=p)\\
        &=P(p(t)=p\,|\,p(t-\epsilon)=p-\D p) \f{\mP(x,p-\D p,t-\epsilon)}{\mP(x,p,t)}\\
        &\approx P(p(t)=p|p(t-\epsilon)=p-\D p)\\
        &\qquad \times\left(1-\D p \partial_p \ln\mP(x,p,t)-\mathcal{O}(\epsilon) \right) ,
    \end{aligned}
\end{equation}
where we Taylor expanded $\mP(x,p-\D p,t-\epsilon)$. Thus, 
\begin{equation}
    \begin{aligned}
    &\avg{p(t)-p(t-\epsilon)|x,p,t} \\
    & = \int_\mathbb{R} d(\D p)\D p\, P(p|p-\D p)\left(1-\D p \partial_p \ln\mP(x,p,t)-\mathcal{O}(\epsilon) \right)\\
    & = (1-\mathcal{O}(\epsilon) )\int_\mathbb{R} d(\D p)\D p\, P(p|p-\D p) \\
    &\quad - \partial_p \ln\mP(x,p,t)\int_\mathbb{R} d(\D p)\D p^2\, P(p|p-\D p)\\
    & =  \exp\left(-\frac{\epsilon}{2}f'\right)\left[(f-gg')\epsilon-  \partial_p \ln\mP(x,p,t)2g^2\epsilon - \mathcal{O}(\epsilon^2) \right],
    \end{aligned}
\end{equation}
where the properties of the gaussian integral have been again applied. Adding all terms and taking the limit:
\begin{equation}
    \begin{aligned}
        \avg{\dot{p}|x,p,t}&=\lim_{\epsilon\rightarrow 0}\f{1}{2\epsilon}\exp\left(-\frac{\epsilon}{2}f'\right)\left[2(f-gg')\epsilon\right. \\
        &\quad \left.- \partial_p \ln\mP(x,p,t)2g^2\epsilon-\mathcal{O}(\epsilon^2)\right]\\
        &=(f-gg')- g^2\partial_p\ln\mP(x,p,t)\\
        &= \frac{1}{\mP}\left[(f-g\p_pg)\mP- g^2 \partial_p \mP\right]=\frac{j_p(x,p,t)}{\mP(x,p,t)},
    \end{aligned}
\end{equation}
where the last equality is given by definition \eqref{eq:FP}, and the proof is completed.

\section{\label{app:positivity} Positivity of entropy production functional}
Recall from the original work by Kullback and Liebler \cite{KL} that the KL-divergence is the functional of probability measures that satisfies \cite{Csiszar}:
\begin{enumerate}
    \item \textbf{Non-negativity}: $D_{KL}[P \| P^\dagger] \geq 0$, consistent with the second law of thermodynamics when interpreted as entropy production, and $D_{KL}[P \| P^\dagger]=0 \iff P=P^\dagger$, consistent with thermodynamic equilibrium.
    \item \textbf{Strong additivity}: ensuring consistency under composition of processes, and yielding extensivity of entropy in the case of independent subsystems.
    \item \textbf{Invariance}: The divergence is invariant under non-singular phase space coordinate transformations (diffeomorphisms), ensuring EP is an objective physical observable.
\end{enumerate}
Among information divergences, the KL-divergence is the unique functional (up to a positive multiplicative constant) that simultaneously satisfies non-negativity, strong additivity (the chain rule), and invariance, once pathological solutions are excluded by mild regularity assumptions (e.g. continuity or measurability). For axiomatic foundations and related characterizations of the KL-divergence, see the Shore-Johnson axioms \cite{SJ_axioms}, and modern treatments \cite{Leinster,Csiszar,TEMPESTA2016}.

Let us now recover the expression for the total EP:
\begin{equation}\label{eq:s_tot_functional}
\begin{aligned}
        \avg{\D s_{\rm tot}} &= \avg{\D s_{sys}}  + \avg{\D \Sigma }= \avg{\D {\mathfrak{s}}_{\rm tot}} - \avg{\D{\mathfrak{s}}_m} + \avg{\D\Sigma }\\
        &= \avg{\D {\mathfrak{s}}_{\rm tot}} -\int_{0 , \o}^{t} d\t\,d\mathbf{u}\:\bm{\mathfrak{j}}\cdot\bm{\phi},
\end{aligned}
\end{equation}
where we have applied $\D s_{sys}= \D {\mathfrak{s}}_{\rm tot} - \D{\mathfrak{s}}_m$ and exchanged the integration order assuming the integrand is absolutely integrable (Fubini's theorem) as $\int_{\mathbf{u}_0}^{\mathbf{u}_t} \mathcal{D}\mathbf{u} P[\mathbf{u}] \int_{0}^{t} d\t \,(\dot\Sigma - \dot{\mathfrak{s}}_m)= \int_{0}^{t} d\t \avgg{\dot\Sigma - \dot{\mathfrak{s}}_m}$. Here, $\bm{\mathfrak{j}}$ denotes the dissipative contributions to the probability currents and $\bm{\phi}(\mathbf{u},\t)$ is a measure of LDB violation $(\text{LDB}\iff  \bm{\phi}\equiv0\iff \dot\Sigma = \dot{\mathfrak{s}}_m)$ independent of $\mP$. This form $\avgg{\dot\Sigma-\dot{\mathfrak{s}}_m }=\int_\Omega d\mathbf{u}\:\bm{\mathfrak{j}}\cdot \bm{\phi}$ is obtained by direct inspection on a case-by-case basis; see Eqs. \eqref{eq:total_EPR_closed}, \eqref{eq:total_EPR_inter}, and \eqref{eq:total_EPR_open}. Note here we assumed a continuous phase space $\o$, but the argument is completely analogous for discrete phase spaces, as well as for any system dimensionality.

Since $\avg{\D \mathfrak{s}_{\rm tot}}:= D_{KL}[P[\mathbf{u}]||P[\hat{\mathbf{u}}]] $ is given as a KL-divergence between forward and backward trajectories $\mathbf{u}:[0,t]\subset\mathbb{R}\rightarrow \o$, it satisfies properties $(1-3)$. The second term satisfies strong additivity (2). This follows by noting that it can be written as $\avg{\D\Sigma -\D{\mathfrak{s}}_m}$, for which strong additivity can be established independently for each contribution.

For $\avg{\D\mathfrak{s}_m }= \int_{\mathbf{u}_0}^{\mathbf{u}_t} \mathcal{D}\mathbf{u} P[\mathbf{u}]\text{ln}\frac{P[\mathbf{u}|\mathbf{u}_0 ]}{P[\hat{\mathbf{u}}|\mathbf{u}_t ]}$, its strong additivity follows from factorization of conditional path measures:
    \begin{widetext}      
    \begin{equation}
    \begin{aligned}
    \avg{\D\mathfrak{s}_m} &= \int \mathcal{D}\mathbf{u}\, P[\mathbf{u}]\ln\frac{P[\mathbf{u}|\mathbf{u}_0]}{P[\hat{\mathbf{u}}|\mathbf{u}_t]} = \int \mathcal{D}\mathbf{u}\, P[\mathbf{u}]\left[\ln\frac{P[\mathbf{u}_A|\mathbf{u}_0]}{P[\hat{\mathbf{u}}_A|\mathbf{u}_{t}]} + \ln\frac{P[\mathbf{u}_B|\mathbf{u}_{A},\mathbf{u}_0]}{P[\hat{\mathbf{u}}_B|\hat{\mathbf{u}}_A,\mathbf{u}_t]} \right]\\
    &= \int \mathcal{D}\mathbf{u}_A\, P[\mathbf{u}_A]\ln\frac{P[\mathbf{u}_A|\mathbf{u}_0]}{P[\hat{\mathbf{u}}_A|\mathbf{u}_{t}]} + \Bavg{\int \mathcal{D}\mathbf{u}_B\, P[\mathbf{u}_B|\mathbf{u}_{A}] \ln\frac{P[\mathbf{u}_B|\mathbf{u}_{A},\mathbf{u}_0]}{P[\hat{\mathbf{u}}_B|\hat{\mathbf{u}}_A,\mathbf{u}_t]}}_{\mathbf{u}_A},
    \end{aligned}
    \end{equation}
        \end{widetext}
    where Bayes theorem $P[\mathbf{u}_A,\mathbf{u}_{B}|\mathbf{u}_0]=P[\mathbf{u}_{A}|\mathbf{u}_0]P[\mathbf{u}_B|\mathbf{u}_{A},\mathbf{u}_0]$ and normalization of conditional probabilities have been applied \cite{Kampen,KL}.

Notably, the log ratio of path weights immediately implies invariance. Defining the Jacobian $\mathcal{M}(\mathbf{z})=|\det (\p \mathbf{u}/\p \mathbf{z})|$ \cite{KL}:
\begin{equation}
\begin{aligned}
\avg{\D\mathfrak{s}_m} &= \int \mathcal{D}\mathbf{u}\, P[\mathbf{u}]\ln\frac{P[\mathbf{u}|\mathbf{u}_0]}{P[\hat{\mathbf{u}}|\mathbf{u}_t]}
 \\& = \int \mathcal{D}\mathbf{z}\,\mathcal{M}(\mathbf{z}) \mathcal{M}^{-1}(\mathbf{z})P[\mathbf{z}]\ln\frac{P[\mathbf{z}|\mathbf{z}_0]\mathcal{M}^{-1}(\mathbf{z})}{P[\hat{\mathbf{z}}|\mathbf{z}_t]\mathcal{M}^{-1}(\mathbf{z})} \\
&=   \int \mathcal{D}\mathbf{z}\, P[\mathbf{z}]\ln\frac{P[\mathbf{z}|\mathbf{z}_0]}{P[\hat{\mathbf{z}}|\mathbf{z}_t]}=\avg{\D\mathfrak{s}_m} ,
\end{aligned}
\end{equation}
which is also reflected by the scalar product in Eq. \eqref{eq:s_tot_functional}. 

The same scalar product structure ensures the invariance of $\avg{\D\Sigma}$, arising due to its construction as an invariant in the nonholonomic constraints \eqref{eq:nonhol_constraints}. Importantly, both $\avg{\D\Sigma}$ and $\avg{\D\mathfrak{s}_m}$ admit the same generic representation in terms of probability currents, i.e. $\int_{0 , \o}^{t} d\t\,d\mathbf{u}\:\bm{\mathfrak{j}}\cdot\bm{\psi}$, where $\bm{\psi}(\o)$ is a state-dependent function independent of the probability measure $\mP$. This shared structure suggests that strong additivity extends to $\avg{\D\Sigma}$ as well, a fact that can be verified explicitly. In this representation, strong additivity leads to a splitting into marginal and conditional terms in the probability measure $\mP$, as opposed to the path weight $P$, since the time integration has been exchanged.

For FP dynamics of the form $\p_t\mP=-\p_\nu\mathfrak{j}^\nu$ with
\begin{equation}
    \mathfrak{j}^\nu=\mathfrak{j}^\nu_{\rm drift}-\mathfrak{j}^\nu_{\rm diff}=F^\nu_{\rm drift} \mP - D^{\nu\mu} \mP\p_\mu\ln\mP ,
\end{equation}
this property is already explicit at the level of the diffusive contribution:
\begin{widetext}    
\begin{equation}
    \begin{aligned}
        \int dxdy\, \mathfrak{j}^\nu_{\rm diff}(x,y)\psi_\nu(x,y) &= \int dxdy\,  D^{\nu\mu} \mP(x,y)\p_\mu\ln\mP(x,y)\psi_\nu(x,y)\\
        & =\int dxdy\,  D^{\nu\mu}  \psi_\nu(x,y)\mP(x)\mP(y|x)\p_\mu\ln\mP(x) + \int dxdy\,  D^{\nu\mu}  \psi_\nu(x,y)\mP(x)\mP(y|x)\p_\mu\ln\mP(y|x)\\
        & =\int dx\,  D^{\nu\mu} \mP(x)\p_\mu\ln\mP(x)\avg{\psi_\nu(x,y)}_{y|x} + \Bavg{\int dy\,  D^{\nu\mu}  \mP(y|x)\p_\mu\ln\mP(y|x) \psi_\nu(x,y)}_x\\
        & =\int dx\,  \mathfrak{j}^\nu_{\rm diff}(x)\avg{\psi_\nu(x,y)}_{y|x} + \Bavg{\int dy\,  \mathfrak{j}^\nu_{\rm diff}(y|x)\psi_\nu(x,y)}_x .
    \end{aligned}
\end{equation}
The drift contribution can be absorbed into either the marginal or the conditional part, as
\begin{equation}
\begin{aligned}
        \int dxdy\, \mathfrak{j}^\nu_{\rm drift}(x,y)\psi_\nu(x,y)&= \int dxdy\, F^\nu_{\rm drift}(x,y)\mP(x)\mP(y|x)\psi_\nu(x,y)\\
        &= \int dx\,  \mP(x)\avg{F^\nu_{\rm drift}(x,y)\psi_\nu(x,y)}_{y|x}=\Bavg{\int dy\,  \mathfrak{j}^\nu_{\rm drift}(y|x)\psi_\nu(x,y)}_x .
\end{aligned}
\end{equation}
Consequently, the drift term plays no role in testing strong additivity; the latter is entirely controlled by the diffusive current.

An analogous decomposition holds for Master equations, where the joint probability evolves as
\begin{equation}
    d_t\mP(x,y)=\sum_{x',y'}\mathfrak{j}^{y,y'}_{x,x'}=\sum_{x',y'}[w^{y,y'}_{x,x'}\mP(x',y')-w^{y',y}_{x',x}\mP(x,y)],
\end{equation}
and $w^{y,y'}_{x,x'}$ denotes the transition rate $(x',y')\to(x,y)$. Using Bayes rule and the identity $\sum_{x',y'}\mathfrak{j}^{y,y'}_{x,x'}=\sum_{x'}\mathfrak{j}^{y}_{x,x'}+\sum_{y'}\mathfrak{j}^{y,y'}_{x}$ \cite{Horowitz_esposito}, one finds:
\begin{equation}
    \begin{aligned}
        \sum_{x,y}\sum_{x',y'}&\mathfrak{j}^{y,y'}_{x,x'}\psi^{y,y'}_{x,x'} =\sum_{x,y}\sum_{x'}\mathfrak{j}^{y}_{x,x'}\psi^{y}_{x,x'}+\sum_{x,y}\sum_{y'}\mathfrak{j}^{y,y'}_{x}\psi^{y,y'}_{x}
\\&= \sum_{x,y}\sum_{x'}[w^{y}_{x,x'}\mP(x',y)-w^{y}_{x',x}\mP(x,y)]\psi^{y}_{x,x'}+\sum_{x,y}\sum_{y'}[w^{y,y'}_{x}\mP(x,y')-w^{y',y}_{x}\mP(x,y)]\psi^{y,y'}_{x}\\
&= \sum_{x,y}\sum_{x'}[w^{y}_{x,x'}\mP(x')\mP(y|x')-w^{y}_{x',x}\mP(x)\mP(y|x)]\psi^{y}_{x,x'}+\sum_{x,y}\sum_{y'}[w^{y,y'}_{x}\mP(x)\mP(y'|x)-w^{y',y}_{x}\mP(x)\mP(y|x)]\psi^{y,y'}_{x}\\
&= \sum_{x,x'}[\mP(x')\avg{w^{y}_{x,x'}\psi^{y}_{x,x'}}_{y|x'}-\mP(x)\avg{w^{y}_{x',x}\psi^{y}_{x,x'}}_{y|x}]+\Bavg{\sum_{y,y'}[w^{y,y'}_{x}\mP(y'|x)-w^{y',y}_{x}\mP(y|x)]\psi^{y,y'}_{x}}_x .
\end{aligned}
\end{equation}
\end{widetext}
This decomposition again separates marginal and conditional contributions, in direct analogy with the FP case. For a detailed discussion of conditional corrections to EPRs see Refs.~\cite{Horowitz_esposito,Allahverdyan_2009}. As this decomposition holds for arbitrary $\bm{\psi}$, it also applies to $\avg{\D \Sigma}$ as a special case; see Eqs. \eqref{eq:med_EPR_closed}, \eqref{eq:med_EPR_inter}, and \eqref{eq:med_EPR_open}. Entropy pumping terms $\D\mI$ cancel between $\D\mathfrak{s}_m$ and $\D \Sigma$ and thus do not contribute to the analysis of $\avg{\D\Sigma -\D{\mathfrak{s}}_m}$.

Thus, altogether, $\avg{\D s_{\rm tot}}$ satisfies properties $(2-3)$. By enforcing the second law (non-negativity) of $\avg{\D s_{\rm tot}}$, the functional is brought into full alignment with the properties $(1-3)$. Given that the KL-divergence is the unique functional satisfying the criteria of invariance, strong additivity, and non-negativity (again under mild regularity), it follows that \begin{equation}
    \avg{\D s_{\rm tot}}\propto \l D_{KL},\qquad \l \geq 0. 
\end{equation}  
The first contribution in Eq. \eqref{eq:s_tot_functional} satisfies this requirement by construction. By contrast, the second contribution contains a log ratio of conditional path weights $(\avg{\D\mathfrak{s}_m})$ together with an additional term that cannot be derived from path probabilities $(\avg{\D\Sigma})$ \cite{Shargel_2010}; therefore, it does not define a KL-divergence. Consequently, imposing the second law requires $\avg{\D\Sigma -\D{\mathfrak{s}}_m}=0\Rightarrow \D\Sigma =\D{\mathfrak{s}}_m$. Since  $\bm{\mathfrak{j}} \equiv 0$ cannot be enforced, this requirement enforces $\bm{\phi} \equiv 0$, thereby recovering LDB as a structural requirement of the second law axiom.

\section{\label{app:mech_closed} Mechanical isolated system derivations}
In this section the detailed derivations of Sec. \ref{sec:mech_closed} are provided. First, we derive the EoMs \eqref{eq:EoM_mech_closed}. The variational principle in this case is of the form:
\begin{equation}\label{equation_q_s}
    \begin{aligned}
    &\delta \int_{t_1}^{t_2} dt\left( L (x,v,s) + p \cdot ( \dot x - v)  + \dot \Gamma(s-\Sigma)\right) =0 , \\ 
    &\begin{cases}
   \vspace{0.2cm}\displaystyle\frac{\partial L}{\partial s}\dot \Sigma= ({\rm f}_-+ g\circ\zeta) \cdot \dot x, \\
   \vspace{0.2cm}\displaystyle\frac{\partial L}{\partial s}\delta \Sigma= ({\rm f}_-+ g\circ\zeta) \cdot\delta x.
\end{cases}
\end{aligned}
\end{equation}
Taking variations of the Lagrangian and taking into account the fixed endpoints condition gives
\begin{align}
    \int_{t_1}^{t_2}dt\left( \f{\p L}{\p x}\d x +\f{\p L}{\p v}\d v +\f{\p L}{\p s}\d s +\d p(\dot x -v) - \dot p \d x - p\d v \nonumber \right.\\
    -  \d\Gamma(\dot s - \dot \Sigma) + \dot \Gamma(\d s - \d \Sigma)\bigg)=0,
\end{align}
such that, applying the nonholonomic constraints,
\begin{subequations}
    \begin{eqnarray}
        && \d p:v=\dot x,\\
        && \d v: p = \f{\p L}{\p v} = mv,\\
        && \d \Gamma:\dot s = \dot \Sigma,\\
       &&  \d s:\dot \Gamma = -\f{\p L}{\p s} = T,\\
        && \d x: \dot p = \f{\p L}{\p x} + {\rm f}_-+ g\circ\zeta,
    \end{eqnarray}
\end{subequations}
which finally gives the system \eqref{eq:EoM_mech_closed}.

To this the system corresponds the FP equation $ \partial_t \mP = -\nabla_\Omega \cdot \mathbf{j}$ with currents
\begin{subequations}
\begin{align}
     j_x =& \mP\f{p}{m},\\
     j_p =& \left(-\partial_x U-\gamma\f{p}{m}-g\partial_p g+\f{p}{m}g\partial_s (g/T)\right)\mP \nonumber\\
    & -g^2\partial_p \mP+\f{pg^2}{mT}\partial_s \mP, \\
    j_s =&  -\f{p}{mT}\jdd .
\end{align}
\end{subequations}
We denote the reversible and irreversible currents as $\jrr = -\partial_x U \mP $ and $\jdd = j_p - \jrr$, respectively.  The average system EPR gives, applying natural boundary conditions \eqref{eq:natural_BC}:
\begin{equation}
\begin{aligned}
    \dot \mS&=-\f{d}{dt}\int d\Omega\,  \mP \ln \mP =-\int d\Omega\,  \ln \mP \partial_t \mP \\ 
    & =  \int d\Omega\,  \ln \mP\nabla_\Omega\cdot \mathbf{j} = -\int d\Omega\,  \f{\mathbf{j}}{\mP}\cdot\nabla_\Omega\mP \\
    &=-\int d\Omega  \left[\f{p}{m}\partial_x \mP -\partial_x U\partial_p \mP +\f{\jdd}{\mP}(\partial_p\mP-\f{p}{mT}\partial_s\mP)\right]\\
    & =\int d\Omega\left[ \mP \left(\partial_x\f{p}{m} -\partial_p\partial_x U \right) -\f{\jdd}{\mP}(\partial_p\mP-\f{p}{mT}\partial_s\mP)\right]\\
    &= \int d\Omega\, \f{\jdd}{\mP g^2} \left[\jdd -\left(-\gamma\f{p}{m}-g\partial_p g+\f{p}{m}g\partial_s (g/T)\right)\mP\right].
\end{aligned}
\end{equation}
The average thermodynamic EPR can be derived by either applying the two-step average $\avgg{\dot s} = \int d\Omega \,j_s$ or from the ensemble description:
\begin{align}
    \f{d\avg{s}}{dt} &= \int d\Omega\, s \partial_t \mP = -\int d\Omega\, s \nabla_\Omega\cdot\mathbf{j} = \int d\Omega\, \mathbf{j}\cdot \nabla_\Omega s \nonumber\\
    &= \int d\Omega\, j_s=-\int d\Omega\, \f{p}{mT}\jdd =\avgg{\dot s}.
\end{align}
Adding both contributions then yields the total EPR given in Eq. \eqref{eq:total_EPR_closed}.

\section{\label{app:pseudo_inv} LDB and Master FT for closed systems}
The covariance matrix of Eqs. \eqref{eq:EoM_mech_closed} corresponds to
\begin{equation}
    D^{\nu\mu}=\begin{bmatrix}
    g^2 &\displaystyle -\f{pg^2}{mT}\\
    \displaystyle -\f{pg^2}{mT} & \displaystyle \left(\f{pg}{mT}\right)^2
\end{bmatrix},
\end{equation}
such that $\det D^{\nu\mu}=0$ and hence the inverse does not exist. However, we can derive the Moore–Penrose pseudo-inverse \cite{meyer01}
\begin{equation}
    D_{\mu\nu}^\dagger =\f{1}{g^2(T^2+(p/m)^2)^2}\begin{bmatrix}
  \vspace{1.5mm}  T^4 &\displaystyle -T^3 \f{p}{m}\\
    \displaystyle-T^3 \f{p}{m} & \displaystyle \left(T\f{p}{m}\right)^2
\end{bmatrix},
\end{equation}
and apply it in the OM action to compute the medium EPR---see Ref. \cite{Ayanbayev_2021} for a rigorous treatment of OM functionals involving pseudo-inverses in Gaussian measures. We denote the even and odd contributions
\begin{subequations}
\begin{align}
    \bf u_+ &= \begin{bmatrix}
    \dot p +\partial_x U \\
     \displaystyle\f{p}{mT}(-\gamma\f{p}{m}-\nu_p)
\end{bmatrix}, \\
    \bf u_- &= \begin{bmatrix}
    \displaystyle\gamma p/m+\nu_p \\ \dot s
\end{bmatrix},
\end{align}
\end{subequations}
with $\nu_p = g\partial_p g-\f{p}{m}g\partial_s (g/T)$. Decomposing $D_{\mu\nu}^\dagger$ into the symmetric and anti-symmetric contributions $D_{\mu\nu}^{\dagger,+} ,D_{\mu\nu}^{\dagger,-}$, respectively, the OM action and its time reversal are given by 
\begin{subequations}
\begin{align}
     \mathcal{A}[\mathbf{u}] =& \int_{t_1}^{t_2} d\tau \bigg(\f{1}{4}(\zrr+\zii)^T D_{\mu\nu}^\dagger (\zrr+\zii) \nonumber\\
        & +\f{1}{2}(\p_p F^p + \p_s F^s)\bigg),\\
      \hat{\mathcal{A}}[\mathbf{u}] =& \int_{t_1}^{t_2}  d\tau \bigg(\f{1}{4}(\zrr-\zii)^T (D_{\mu\nu}^{\dagger,+}  - D_{\mu\nu}^{\dagger,-} )(\zrr-\zii) \nonumber\\
      & +\f{1}{2}(-\p_p \hat F^p + \p_s \hat F^s)\bigg).
\end{align}
\end{subequations}
The log ratio \eqref{eq:EP} then gives (recall the measure contribution cancels):
\begin{align}
     \hat{\mathcal{A}}[\mathbf{u}]- \mathcal{A}[\mathbf{u}] =& \int_{t_1}^{t_2} d\tau\bigg(\f{1}{4}\left( -2\zrr D_{\mu\nu}^{\dagger,-} \zrr-2\zii D_{\mu\nu}^{\dagger,-} \zii \right. \nonumber\\
      &-\left.4\zrr D_{\mu\nu}^{\dagger,+}  \zii\right) -\p_pF_+^p - \p_s F_-^s  \bigg),
\end{align}
which after some algebra simplifies to
\begin{align}
    \D \mathfrak{s}_m = \int_{t_1}^{t_2} d\tau \,\f{1 }{g^2}\left( -\gamma\f{p}{m}+g\circ\zeta\right)\left(-\gamma\f{p}{m}-\nu_p \right)  .
\end{align}
Differentiating and enforcing the FDR \eqref{eq:FDR_mech_closed}, we find
\begin{align}
    \dot{\mathfrak{s}}_m  &= \f{d}{d\tau}\text{ln}\frac{P[\mathbf{u}|\mathbf{u}_0 ]}{P[\hat{\mathbf{u}}|\mathbf{u}_t ]} = -\f{1}{T}\left( -\gamma\f{p}{m}+g\circ\zeta\right)\cdot \f{p}{m} \nonumber\\
    &= \dot \Sigma.
\end{align}
The last equality holds by definition, as $\dot s = \dot \Sigma$ for closed systems; see Eqs. \eqref{eq:EoM_mech_closed}. Consequently, $\D \Sigma$ is consistent with the log ratio of path probabilities under the constraint of the FDR, and therefore satisfies both LDB [Eq. \eqref{eq:LDB_VP}] and the master FT \eqref{eq:masterFT}, as established by the argument presented in Sec. \ref{sec:mech_open}. The same result holds upon inclusion of an external manipulation ${\rm f}_+$, demonstrating that the master FT remains valid for both closed and open systems, including those influenced by external irreversible processes or drivings.

\section{\label{app:inter} Interconnected systems derivations}
In this section the detailed derivations of Sec. \ref{sec:inter} are provided. Firstly, taking variations of the Lagrangian and taking into account the fixed endpoints condition gives
\begin{align}
    \int_{t_1}^{t_2}dt\sum_i\left( \f{\p L}{\p N_i}\d N_i +\f{\p L}{\p s_i}\d s_i+\dot W^i\d N_i - \dot N_i\d W^i \nonumber \right.\\
    -  \d\Gamma^i(\dot s_i - \dot \Sigma_i) + \dot \Gamma^i(\d s_i - \d \Sigma_i)\bigg)=0.
\end{align}
Applying the nonholonomic constraints,
\begin{subequations}
    \begin{align}
        & \d N_i: -\f{\p L}{\p N_i}=\dot W^i =\mu^i,\\
        &  \d s_i:-\f{\p L}{\p s_i} =\dot \Gamma ^i= T^i,\\
        & \d \Gamma^i:\dot s_i = \dot \Sigma_i,\\
        & \d W^i: \dot N_i=(\mJ_{j i}+\s_{j i}\circ\zeta),
    \end{align}
\end{subequations}
which finally gives the system \eqref{eq:EoM_interc}. To this system corresponds the FP with probability currents given by Eqs. \eqref{eq:currents_inter}. The full expression of the mass and heat fluxes including the noise-induced drifts $\nu_m ,\nu_{th}$ are  
\begin{widetext}
\begin{align}
    &\tilde{\mJ}_{12} = \mJ_{12}+\s_{12}\p_{N_1}\s_{12} -\s_{12}\p_{N_2}\s_{12} -\s_{12}\p_{s_1}\left(\f{T^1-T^2}{T^1}C\kappa_{12}+\f{\mu^1}{T^1}\s_{12}\right) -\s_{12}\p_{s_2}\left(\f{T^2-T^1}{T^2}C\kappa_{12} - \f{\mu^2}{T^2}\s_{12}\right), \\
    &\tilde{J}_{12} = J_{12} +\kappa_{12}\p_{N_1} C\s_{12} - \kappa_{12}\p_{N_2}C\s_{12} - \kappa_{12}\p_{s_1}\left(\f{T^1-T^2}{T^1}\kappa_{12}+\f{\mu^1}{T^1}C\s_{12}\right) - \kappa_{12}\p_{s_2}\left(\f{T^2-T^1}{T^2}\kappa_{12} - \f{\mu^2}{T^2}C\s_{12}\right).
\end{align}
\end{widetext}

We now derive the total EPR. Firstly, the average thermodynamic EPR can be derived by either applying the two-step average or from the ensemble description, recalling  the natural boundary conditions \eqref{eq:natural_BC}:
\begin{align}
    & \f{d}{dt}\sum_k\avg{s_k} = \int d\Omega\, (s_1+s_2) \partial_t \mP = -\int d\Omega\,  (s_1+s_2)  \nabla_\Omega\cdot\mathbf{j} \nonumber\\ 
    &= \int d\Omega\, \mathbf{j}\cdot \nabla_\Omega  (s_1+s_2)  = \int d\Omega\, (j_{s_1}+j_{s_2}) =\sum_k\avgg{\dot s_k} \nonumber\\
    &= \int d\Omega\left\{j_{N_1}\left(\frac{\mu^2}{T^2}-\frac{\mu^1}{T^1}\right)+j_{th}\left(\f{T^1-T^2}{T^1}+\f{T^2-T^1}{T^2}\right)  \right\}.
\end{align}
Then, the system EPR is computed as follows,
\begin{widetext}
    \begin{align}
    \dot \mS&=-\f{d}{dt}\int d\Omega\,  \mP \ln \mP =-\int d\Omega\,  \ln \mP \partial_t \mP  =  \int d\Omega\,  \ln \mP\nabla_\Omega\cdot \mathbf{j} = -\int d\Omega\,  \f{\mathbf{j}}{\mP}\cdot\nabla_\Omega\mP \nonumber\\
    &=-\int \f{d\Omega}{\mP}  \left\{j_{N_1}(\p_{N_1}\mP-\p_{N_2}\mP)+j_{N_1}\left(\f{\mu^2}{T^2}\p_{s_2}\mP-\f{\mu^1}{T^1}\p_{s_1}\mP\right)+j_{th}\left(\f{T^1-T^2}{T^1}\p_{s_1}\mP+\f{T^2-T^1}{T^2}\p_{s_2}\mP\right)\right\} \nonumber\\
    &=-\int \f{d\Omega}{\mP}  \left\{\f{j_{N_1}}{\s^2}\left[-\tilde{\mJ}_{12}\mP-j_{N_1}+C\kappa\s\left(\f{T^1-T^2}{T^1}\p_{s_1}\mP+\f{T^2-T^1}{T^2}\p_{s_2}\mP\right)\right]+\f{j_{th}}{\kappa^2}\left[\tilde{J}_{12}\mP-j_{th}-\f{C\kappa}{\s}(\tilde{\mJ}_{12}\mP+j_{N_1})\right]\right\} \nonumber\\
    &=-\int \f{d\Omega}{\mP(1-C^2)}  \left\{\f{j_{N_1}}{\s^2}\left[\left(-\tilde{\mJ}_{12}+\f{C\s}{\kappa}\tilde{J}_{12}\right)\mP-j_{N_1}-\f{C\s}{\kappa}j_{th}\right]+\f{j_{th}}{\kappa^2}\left[\left(\tilde{J}_{12}-\f{C\kappa}{\s}\tilde{\mJ}_{12}\right)\mP-j_{th}-\f{C\kappa}{\s}j_{N_1}\right]\right\} \nonumber\\
    &=\int \f{d\Omega}{(1-C^2)}\left\{ \f{j^2_{N_1}}{\mP\s^2 } + \f{j^2_{th}}{\mP\kappa^2} +2C\f{j_{N_1}j_{th}}{\mP\s\kappa} +j_{N_1}\left(\f{\tilde{\mJ}_{12}}{\s^2}-\f{C}{\kappa\s}\tilde{J}_{12}\right) - j_{th}\left(\f{\tilde{J}_{12}}{\kappa^2}-\f{C}{\kappa\s}\tilde{\mJ}_{12}\right) \right\},
\end{align}
\end{widetext}
where we used Eqs. (\ref{eq:currents_inter},~\ref{eq:j_th_inter}). Note in Sec. \ref{sec:inter} $\avgg{\dot\Sigma},\dot\mS$ are rewritten in terms of $j_{N_2}=-j_{N_1}$ for sign convenience.

For completeness, the identification of $\avgg{\dot{\mathfrak{s}}_m }$ in Eq. \eqref{eq:S_sys_inter} is verified via the PI approach. Although a brute force computation applying the Moore–Penrose pseudo-inverse to the system \eqref{eq:EoM_interc} gives the desired result, as demonstrated in Appendix \ref{app:pseudo_inv}, here we follow the approach in Ref. \cite{Cates_2022}, where the system \eqref{eq:EoM_interc} is rewritten as:
\begin{equation}
    \begin{aligned}
       & \dot N_1 = -\mathfrak{j}_N ,\\
       & \dot N_2 =\mathfrak{j}_N ,\\
        & T^1\dot s_1=\mathfrak{j}_S (T^1-T^2)+\mathfrak{j}_N\mu^1 ,\\
       &  T^2 \dot s_2=\mathfrak{j}_S(T^2-T^1)-\mathfrak{j}_N\mu^2, \\
       & \mathfrak{j}_N = \mJ_{12}+\s_{12}\circ\zeta, \\
       &  \mathfrak{j}_S  = J_{12}+\kappa_{12}\circ\eta.
    \end{aligned}
\end{equation}
Then, the PI action can be reduced to the equations for $\mathfrak{j}_N, \mathfrak{j}_S$ which give the non-degenerate covariance matrix $\mathbb{L}$. Note this structure is naturally revealed by the Shannon entropy change in Eq. \eqref{eq:S_sys_inter}. The forward and backward actions then take the form:
\begin{align}
    & \mathcal{A}[\{N_i\},\{s_i\}] = \int_{t_1}^{t_2} \f{d\tau}{4}\begin{bmatrix}
        \mathfrak{j}_N - \tilde{\mJ}_{12} \\  \mathfrak{j}_S  - \tilde{J}_{12}
    \end{bmatrix} \mathbb{L}^{-1}\begin{bmatrix}
        \mathfrak{j}_N - \tilde{\mJ}_{12} \\  \mathfrak{j}_S  - \tilde{J}_{12}
    \end{bmatrix}^T,\\
     & \hat{\mathcal{A}}[\{N_i\},\{s_i\}] = \int_{t_1}^{t_2} \f{d\tau}{4}\begin{bmatrix}
        -\mathfrak{j}_N - \tilde{\mJ}_{12} \\ - \mathfrak{j}_S  - \tilde{J}_{12}
    \end{bmatrix} \mathbb{L}^{-1}\begin{bmatrix}
       - \mathfrak{j}_N - \tilde{\mJ}_{12} \\  -\mathfrak{j}_S  - \tilde{J}_{12}
    \end{bmatrix}^T,
\end{align}
where $\mathfrak{j}_N, \mathfrak{j}_S$ are odd under time-reversal (see \cite{Cates_2022}). After some algebra it is found:
\begin{equation}
      \D \mathfrak{s}_m =  \hat{\mathcal{A}} - \mathcal{A}  = \int_{t_1}^{t_2} d\tau\begin{bmatrix}
         \tilde{\mJ}_{12} \\  \tilde{J}_{12}
    \end{bmatrix} \mathbb{L}^{-1}\begin{bmatrix}
        \mJ_{12}+\s_{12}\circ\zeta\\   J_{12}+\kappa_{12}\circ\eta 
    \end{bmatrix}^T.
\end{equation}
Taking the time derivative followed by the two-step average yields:
\begin{equation}
   \avgg{\dot{\mathfrak{s}}_m} = \int d\Omega \begin{bmatrix}
        j_{N_2} \\ j_{th}
    \end{bmatrix}^T  \mathbb{L}^{-1} \begin{bmatrix}
        \tilde{\mJ}_{12} \\ \tilde{J}_{12}
    \end{bmatrix} ,
\end{equation}
in agreement with Eq. \eqref{eq:S_sys_inter}.

\section{\label{app:mech_open} Mechanical open system derivations}
In this section the detailed derivations of Sec. \ref{sec:mech_open} are provided. First, we derive the EoMs \eqref{eq:EoM_open}. The variational principle in this case is of the form:
\begin{equation}
    \begin{aligned}
    &\delta \int_{t_1}^{t_2}dt\left( L (x,v,s) + p  \cdot ( \dot x - v)  + \dot \Gamma(s-\Sigma)\right)  \\
     &\hspace{5cm}  +\int_{t_1}^{t_2}dt\; {\rm f}_+\cdot \delta x =0, \\ 
    &\begin{cases}
   \vspace{0.2cm}\displaystyle\frac{\partial L}{\partial s}\dot \Sigma= (f_-+ g\circ \zeta) \cdot\dot x+(\mathcal{J}_{s,h}+\k\circ\eta)\cdot(\dot \Gamma-T^h)\\
   \hfill +\mI_p\, \dot\Gamma, \\
   \vspace{0.2cm}\displaystyle\frac{\partial L}{\partial s}\delta \Sigma=  (f_-+ g\circ \zeta) \cdot\delta x+(\mathcal{J}_{s,h}+\k\circ\eta)\cdot\delta\Gamma \\
   \hfill +\mI_p\, \d\Gamma.
\end{cases}
\end{aligned}
\end{equation}  
Taking variations of the Lagrangian and taking into account the fixed endpoints condition gives
\begin{align}
    \int_{t_1}^{t_2}dt\left( \f{\p L}{\p x}\d x +\f{\p L}{\p v}\d v +\f{\p L}{\p s}\d s +\d p(\dot x -v) - \dot p \d x - p\d v \nonumber \right.\\
    -  \d\Gamma(\dot s - \dot \Sigma) + \dot \Gamma(\d s - \d \Sigma) + {\rm f}_+\cdot \delta x\bigg)=0,
\end{align}
such that, applying the nonholonomic constraints,
\begin{subequations}
    \begin{eqnarray}
        && \d p:v=\dot x,\\
        && \d v: p = \f{\p L}{\p v} = mv,\\
       &&  \d s:\dot \Gamma = -\f{\p L}{\p s} = T,\\
        && \d \Gamma:\dot s - \dot \Sigma = \mathcal{J}_{s,h}+\k\circ\eta +\mI_p,\\
        && \d x: \dot p = \f{\p L}{\p x} + {\rm f}_+ + {\rm f}_-+ g\circ\zeta,
    \end{eqnarray}
\end{subequations}
which finally gives the system \eqref{eq:EoM_open}.

To this the system corresponds the FP equation $ \partial_t \mP = -\nabla_\Omega\cdot \mathbf{j}$ with currents
\begin{subequations}\label{eq:FPE_open}
\begin{align}
     j_x &= \mP\f{p}{m},\\
     j_p &= \left(-\partial_x U+ {\rm f}_+ \right)\mP + j_d, \\
    j_s &=  -\f{p}{mT}\jdd + \f{T^h}{T}j_{th},
\end{align}
\end{subequations}
with $j_d ,j_{th}$ as defined in Eqs. \eqref{eq:currents_open}.  Since in this case $\dot s \neq \dot \Sigma$, the second law relies on the average medium EPR, see Eq. \eqref{eq:2nd_law_VP}, for which we apply $\avg{\dot{p}|x,p,s,t}=j_p /\mP$ and $\avg{\dot{s}|x,p,s,t}=j_s /\mP$:
\begin{equation}
\begin{aligned}
     \avgg{\dot \Sigma}&=\avgg{\dot s}-\avgg{\dot s_f} \\
     &= \int d\Omega\, \left(-\f{p}{mT}\jdd+ \f{T^h-T}{T}j_{th}\right)-\avg{\p_p {\rm f}_+}.
\end{aligned}
\end{equation}
The average system EPR gives, applying natural boundary conditions \eqref{eq:natural_BC}:
\begin{widetext} 
\begin{equation}
    \begin{aligned}
        \dot \mS &= - \int d\Omega\, \ln \mP\, \f{\p \mP}{\p t}
    =  \int d\Omega\, \ln \mP\,   \nabla_\Omega \cdot \jv=-\int d\Omega\,  \f{\mathbf{j}}{\mP}\cdot\nabla_\Omega\mP\\
  &  =-\int d\Omega\,  \left(\f{p}{m}\partial_x \mP +(-\partial_x U+{\rm f}_+)\partial_p \mP +\f{\jdd}{\mP}(\partial_p\mP-\f{p}{mT}\partial_s\mP)+\f{T^h}{T}\f{j_{th}}{\mP}\p_s\mP\right)\\
    & =\int d\Omega\, \mP \left(\partial_x\f{p}{m} +\partial_p(-\partial_x U+{\rm f}_+) \right) -\int d\Omega\,\f{\jdd}{\mP}(\partial_p\mP-\f{p}{mT}\partial_s\mP)-\int d\Omega\,\f{T^h}{T}\f{j_{th}}{\mP}\p_s\mP\\
    &= \avg{\partial_p ({\rm f}_+ -\partial_x U)}+\int d\Omega\left\{\f{\jdd}{\mP g^2} \left[\jdd -(-\gamma\f{p}{m}-g\partial_p g+\f{p}{m}g\partial_s (g/T))\mP\right]+\f{j_{th}}{\mP\k^2}\left(j_{th}-\left[\mathcal{J}_{s,h}-\k\p_s\left(\f{T^h}{T}\k\right)\right]\mP\right)\right\}\\
    &=\int \f{d\Omega}{\mP}\,\left(\f{\jdd^2}{g^2}+\f{j_{th}^2}{\k^2}\right)-\int d\Omega\, \f{\jdd}{\ g^2} \left(-\gamma\f{p}{m}-g\partial_p g+\f{p}{m}g\partial_s (g/T)\right) -\int d\Omega\,\f{j_{th}}{\k^2}\left(\mathcal{J}_{s,h}-\k\p_s\left(\f{T^h}{T}\k\right)\right) +\avg{\partial_p {\rm f}_+ }.
    \end{aligned}
\end{equation}
\end{widetext}

% \nocite{*}

\bibliography{bibliography}% Produces the bibliography via BibTeX.

\end{document}